\documentclass[preprint,review,12pt]{elsarticle}

\usepackage[margin=2.2cm]{geometry}

\usepackage{amssymb}
\usepackage{amsmath}
\usepackage{amsthm}
\usepackage{lineno}
\usepackage{float}
\usepackage{hyperref}

\journal{Composite Science and Technology}

\begin{document}

\begin{frontmatter}

\title{Autoregressive Modelling and Synthetic Generation of High-Fidelity, Statistically Equivalent 3D Microstructures for As-Manufactured Misalignments in Fiber-Reinforced Composites}

\author[tud]{Mohamad A. Raja\corref{cor1}}
\ead{M.A.Raja@tudelft.nl}

\author[tud]{Clemens Dransfeld}

\author[suda]{Boyang Chen\corref{cor2}}
\ead{boyangchen@suda.edu.cn}

\cortext[cor1]{Corresponding author. Address: Kluyverweg 1, 2629 HS Delft, The Netherlands. Office: Building 62, NB0.42.}
\cortext[cor2]{Corresponding author. Address: No. 1 Shizi Street, 2215006, China. No. 1 Shizi Street}

\affiliation[tud]{organization={Delft University of Technology (TU Delft), Faculty of Aerospace Engineering, Department of Aerospace Structures and Materials},
            addressline={Kluyverweg 1},
            city={Delft},
            postcode={2629 HS},
            country={The Netherlands}}

\affiliation[suda]{organization={Soochow University, School of Optoelectronic Science and Engineering},
            addressline={No. 1 Shizi Street},
            city={Suzhou},
            postcode={215006},
            country={China}}
\begin{abstract}
This study presents an integrated framework for processing, modelling, and generating statistically representative three-dimensional fiber microstructures from experimental X-ray-$\mu$CT observations. First, an analytical slice-segment ellipse-intersection method is introduced to extract per-slice and per-fiber in-plane and out-of-plane misalignment profiles along the fiber depth. These descriptors are then used to construct a stochastic model that captures slice-wise misalignment distributions and their depth-wise evolution through, copula-based in-plane dependence, latent autoregressive continuity, and rare extreme-misalignment motifs. The model hyperparameters are calibrated using Bayesian optimization, achieving close agreement with the original statistical descriptors, with deviations generally below 10\%. The optimized statistical model is coupled with a physical generation strategy that begins with variable-radius fiber seeding layer and proceeds through an iterative slice-by-slice 3D growth scheme, where the statistical layer guides fiber evolution and Delaunay-based neighbourhood construction with ellipse-based contact resolution ensures non-overlapping, radius-augmented synthetic microstructures. The framework successfully generates about 2400 synthetic fibers while preserving strong statistical fidelity to the original X-ray-$\mu$CT data. The proposed pipeline provides a promising and scalable route for generating statistically equivalent, geometrically admissible, and simulation-ready fiber composite microstructures for virtual testing and analysis. 
\end{abstract}

\begin{keyword}
Microstructure generation \sep Fiber reinforced polymers (FRPs) \sep Representative volume elements (RVEs) \sep Statistically equivalent \sep Numerical methods
\end{keyword}

\end{frontmatter}
\newpage


\section{Introduction}
\label{sec1}

Fiber-reinforced polymer composite (FRP) materials have become indispensable in the design and manufacturing of advanced structures and components, such as aircraft structures, with percentages of well over 50\% in the most recent aircraft. This adoption reflects the superior strength-to-weight ratio, high specific stiffness, design flexibility, corrosion resistance, and fatigue performance of composites \cite{ref1}.

Yet, the outstanding macroscopic performance of these materials is ultimately governed not only by the constituent properties \cite{ref2,ref3}, but also by the mechanisms that originate at the microscale, for instance, local deviations in fiber clustering \cite{ref4,ref5}, fiber orientation, waviness, porosity \cite{ref6,ref7,ref8,ref9,ref10}, and the fiber-matrix interface \cite{ref11,ref12}. Among the various sources of variability, fiber misalignment is particularly critical because it couples geometric imperfection with strong anisotropy in load transfer. Small deviations from the intended ``perfectly aligned'' fiber direction can trigger complex local stress states, adversely and non-uniformly affecting the composite compressive strength depending on the microstructure spatial misalignment distribution and fiber undulation \cite{ref13,ref14,ref15,ref16,ref17,ref18,ref19,ref20}.

On that note, recent advances in high-resolution imaging have substantially improved our ability to interrogate such microstructural features directly. Techniques such as micro-computed tomography ($\mu$CT) \cite{ref21,ref22}, synchrotron and X-ray tomography, and complementary microscopy methods can now resolve fiber trajectories and local packing with increasing fidelity, enabling quantitative measurement of orientation distributions, waviness, inter-fiber spacing, and defect populations \cite{ref23,ref24,ref25,ref26,ref27,ref28,ref29,ref30,ref31}. These imaging modalities make it possible to move beyond idealized assumptions by providing microstructure-informed inputs for computational micromechanics and virtual testing, while also enabling failure analysis that supports microstructure-informed process optimization and more controlled manufacturing conditions through microstructure modeling and synthetic generation.

Thus, several efforts have focused on three-dimensional microstructure modelling and generation of composites incorporating fiber misalignments. For instance, Zheng et al. \cite{ref26} used an optimized fiber path reconstruction algorithm applied to X-ray-$\mu$CT micrographs, in which fiber segmentation was implemented using a U-Net deep learning method and a tracking algorithm. Then, statistical fitting distributions, namely normal, lognormal, and Weibull, were used to analyze differential tortuosity, fiber angle, curvature, and wave magnitude. Seon and Makeev \cite{ref32} developed a computer-graphics-based methodology for generating realistic three-dimensional micromodels of unidirectional CFRPs with stochastic fiber misalignment via Blender's built-in engine for collision detection and rigid-body dynamics to avoid interpenetration, with models containing up to 600 non-intersecting fibers ($V_f > 50\%$), where misaligned fibers are generated using an emitter plane and B\'ezier-curve undulations, showing theoretical misalignment distributions with means and standard deviations below $1^\circ$.

A semi-random empirical framework was proposed by others \cite{ref33,ref34}, in which fiber misalignment is represented using B\'ezier curves while minimizing the standard error of the likelihood and probability to match experimental measurements. Initially straight fibers were implemented in an RVE with approximately 80 fibers, then misalignment was introduced through the fibers' control points with a constant contact distance definition of $\leq 2R$. An electrostatic interaction analogy approach \cite{ref35} was also proposed, showing how randomly placed ellipsoids acting as charged obstacles affect the generation of waved fibers by means of electrostatic interaction using a simplified Newtonian approach. In this approach, an initially random sphere, resembling the initial point of the fiber, is given a velocity and unit charge. Then, due to the electrostatic repulsive forces from the ellipsoids, the sphere moves and its path is tracked. This resulted in an RVE with periodic boundary conditions (PBCs) containing 50 fibers.

Moreover, the authors in \cite{ref36} presented a multiscale statistical characterization framework based on fitting approaches of X-CT scanning results to analyze fiber misalignment, in which, at the mesoscopic level, a modified elliptical symmetry angular Gaussian (ESAG) model effectively characterizes the fiber misalignment. At the microscale, local fiber waviness is described via a cosine-series representation. Fibers with high misalignment, labeled as abnormal fibers, were analyzed using the developed mathematical model, and their appearance was confirmed at the interface of different fiber bundles or resin-rich areas. Others \cite{ref37} used a fused sequential addition and migration (fSAM) algorithm to generate hybrid composites with long-fiber reinforcement, with limited control on fiber curvature and a fiber volume fraction of up to 35\%. Others preserved the original fiber locations and used reconstruction, overlap-removal, and meshing techniques to generate misaligned fiber microstructures \cite{ref38}.

Most works consider the modelling process of fiber misalignment through fitting approaches of global angle distributions, with limited means to characterize the microstructures, or through generating non-intersecting fiber microstructures mimicking the original experimental dataset.

In the current work, a comprehensive framework is presented to systematically analyze X-ray-$\mu$CT datasets slice by slice, assembling rich data structures of local misalignments. Then, a stochastic synthetic model is devised via a modified Gaussian copula modelling scheme to capture planar misalignment correlations, while a per-fiber depth memory is applied through an autoregressive model to capture the  fiber's longitudinal direction behavior. The statistically equivalent modelling layer is calibrated via Bayesian optimization, including high-misalignment fibers. Then, a non-overlapping solver based on projected Gauss-Seidel (PGS) is developed to generate fibers based on slice-by-slice online modelling framework, resulting in rich, high-fidelity generated microstructures with statistically equivalent random fiber arrangements analogous to the real microstructure. 

\begin{figure}[H]
    \centering
    \includegraphics[width=\textwidth]{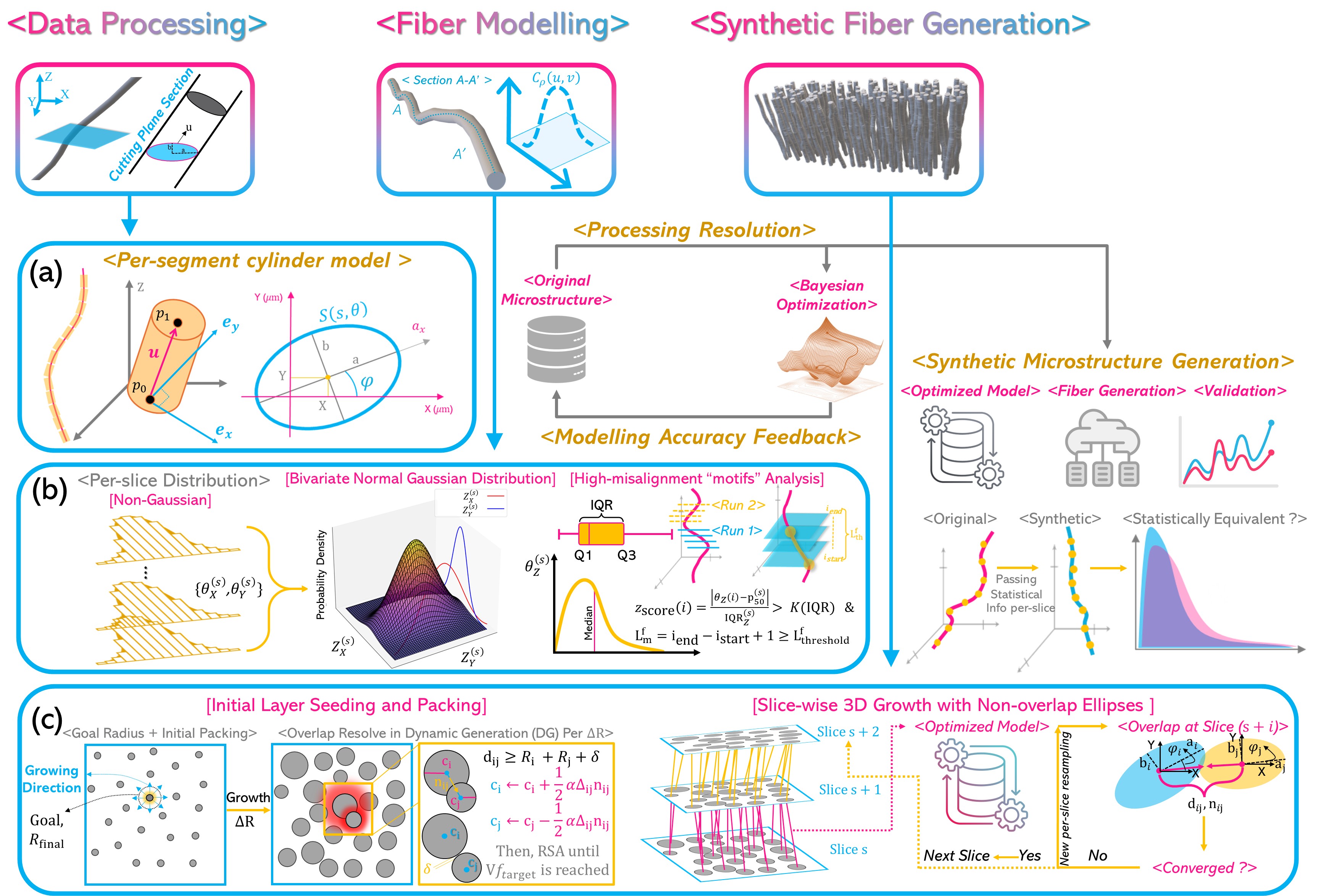}
    \caption{A schematic of the proposed pipeline illustrating the processing, modelling, and synthetic fiber-generation stages. The experimentally observed microstructure is first processed to extract fiber geometry and misalignment descriptors, then statistically modelled using a Bayesian-optimized calibration step, and finally used to generate a statistically equivalent synthetic microstructure. (a) Cylindrical parameterization with an ellipse--plane intersection method to quantify in-plane and along-fiber misalignment. (b) Bivariate Gaussian copula model, including explicit treatment of high-misalignment fibers. (c) Initial layer seeding using dynamic generation (DG) based repulsion and random sequential adsorption (RSA) based insertion, followed by 3D slice-by-slice synthetic fiber generation driven by the online-calibrated model and enforced by a non-overlapping ellipse packing solver enabled by PGS.}
    \label{fig1}
\end{figure}

Thus, the goal of this work is to provide an autonomous virtual-lab experience to seamlessly use the processing, synthetic modelling, and generation pipeline with any type of fiber-based microstructure, in order to generate physically consistent synthetic media that respect the local and global statistics, distributions, and realism of the original microstructure. A schematic of the framework is illustrated in Fig.~\ref{fig1}. The three main sections of this work are as follows: (1) Processing: sequential per-fiber and per-slice fiber misalignment segmentation using an ellipse-intersection method and verification via the central difference method (CDM) as illustrated in Fig.~\ref{fig1}(a). (2) Modelling: development of a stochastic modified Gaussian copula model of synthetic fibers, taking into consideration high-misalignment abnormal fibers, referred to as ``motifs''. This is schematically shown in Fig.~\ref{fig1}(b). The model is optimized using Bayesian optimization against the original microstructure. (3) Generation: development of a physical constraint engine based on an online slice-by-slice optimized model to generate random synthetic microstructures with statistical equivalence to the original microstructure, as shown in Fig.~\ref{fig1}(c).

\section{Methodology}

This section outlines the methodology used to analyze three-dimensional X-ray-$\mu$CT datasets of a unidirectional carbon fiber--reinforced composite prepreg. All computations were performed using open-source libraries together with in-house Python code. Fiber misalignment processing was carried out on an Intel(R) Core(TM) Ultra 5 125U (1.30~GHz) workstation with 16~GB RAM. The Bayesian-optimization-based modelling and the final statistically equivalent microstructure generation were executed on the DelftBlue high-performance computing (HPC) cluster at TU Delft \cite{ref39}. The synthetic fiber microstructure modelling and generation runs used 2 CPUs and 3~GB memory per CPU on Phase-1 compute nodes (48 CPUs and 185~GB RAM per node), equipped with Intel Xeon Gold 6248R processors (24 cores, 3.0~GHz). The Bayesian optimization process required 3.25~hr and the final physical statistical generation required 5.28~hr. 

\subsection{Materials and X-ray-$\mu$CT data acquisition}

The experimental dataset analyzed in this study originates from the work in \cite{ref28}, which reports the X-ray-$\mu$CT acquisition parameters, fiber path reconstruction procedure, and a detailed characterization of the three-dimensional spatial morphology of a thermoplastic CFRP tape at single-fiber resolution. Briefly, a tape segment measuring approximately 2.5~mm in width, orthogonal to the main fiber direction, and 10~mm along the main fiber direction was extracted for microstructural analysis. The segmented volume was subsequently partitioned into six sub-volumes, each with dimensions of approximately $1200~\mu\text{m} \times 160~\mu\text{m} \times 500~\mu\text{m}$ (fiber depth direction). In the present work, the analysis is performed on one of these sub-volumes.

\subsection{Fiber misalignment processing}

Prior to fiber misalignment processing, a clean-up step is required to distinguish continuous from non-continuous fibers. It is important to note that the non-continuous fibers are not physically chopped or inherently short; rather, they are fibers whose tracked depth does not span the full volume due to limitations in imaging and reconstruction. A visual comparison of the original microstructure and the identified continuous and non-continuous fibers is provided in the Supplementary Material (Fig.~S1). The following subsection describes the fiber misalignment frameworks and their derivation applied to the continuous fibers in the sub-volume.

\subsubsection{Ellipse intersection method}

\begin{figure}[H]
    \centering
    \includegraphics[width=\textwidth]{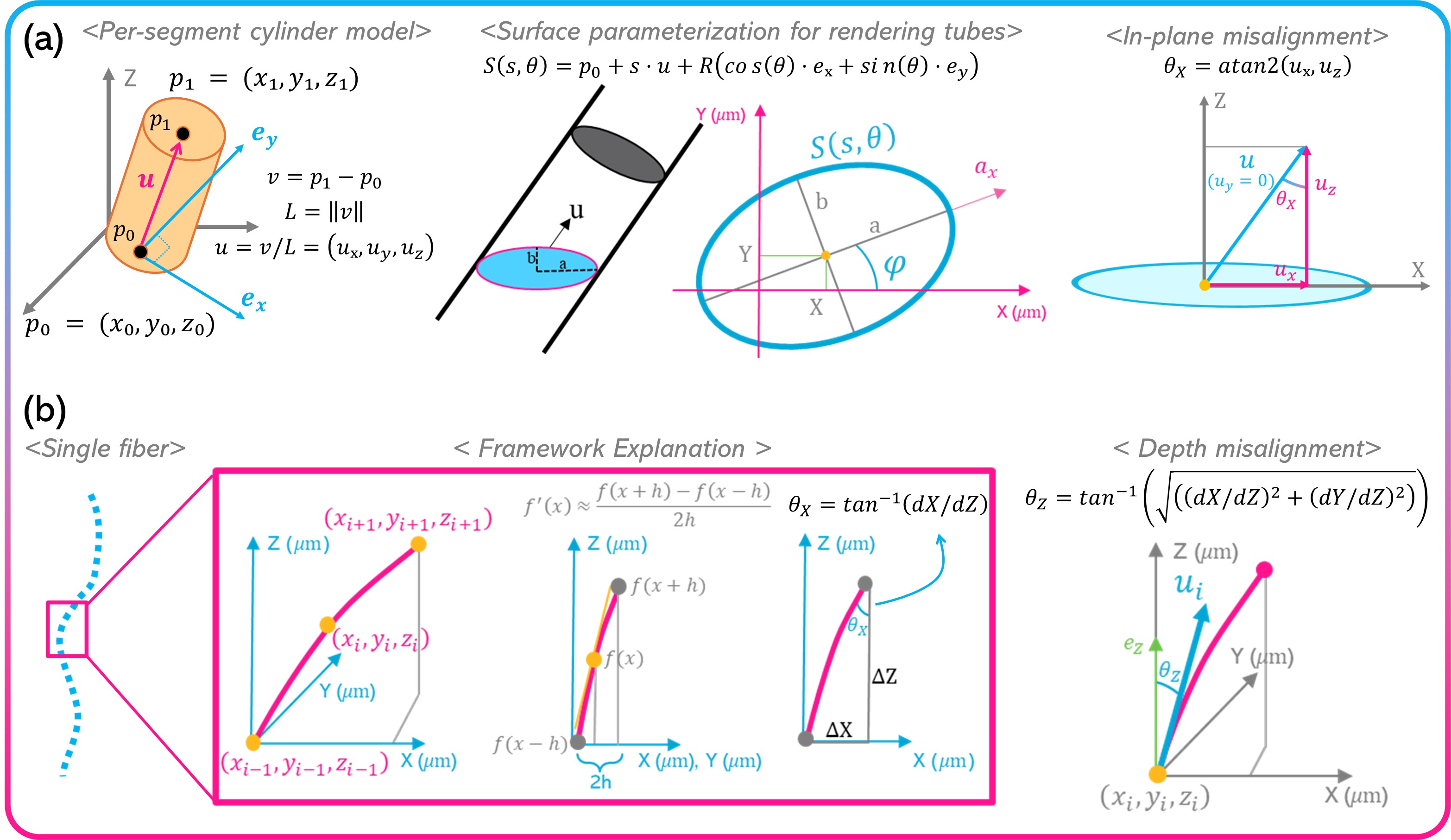}
    \caption{Fiber misalignment quantification in the depth and in-plane directions using (a) the ellipse–intersection method and (b) the central-difference method.}
    \label{fig2}
\end{figure}

This section describes the geometric procedure used to quantify fiber misalignment from three-dimensional X-ray-$\mu$CT centerline data. Each fiber is processed segment-wise by approximating the local trajectory between two consecutive centerline points as a straight cylindrical segment whose axis direction is the local unit tangent vector with radius $R = D/2$, ($D=7$ $\mu$m, only used in the visualization of the intersection,  but does not affect misalignment derivation, however, in physical generation Section~\ref{sec:synthetic_generation} real variable fiber sizes are used). For a given slice plane, $Z = z_{\text{slice}}$, segments that intersect the plane are identified and used to compute (i) the through-thickness inclination $\theta_Z$ (three-dimensional tilt from the global $Z$-axis) and (ii) the in-plane misalignment components $\theta_X$ and $\theta_Y$ (tilts in the $ZX$ and $ZY$ planes). The method avoids numerical differentiation of noisy centreline coordinates and provides a robust per-slice description of misalignment suitable for subsequent statistical modelling. A schematic of the method is presented in Fig.~\ref{fig2}(a).

For two consecutive centerline points, $\mathrm{p}_0 = (x_0, y_0, z_0)$ and $\mathrm{p}_1 = (x_1, y_1, z_1)$, the vector between the points is defined as follows:
\begin{equation}
\mathrm{v} = \mathrm{p}_1 - \mathrm{p}_0, \qquad
L = \lVert \mathrm{v} \rVert, \qquad
\mathrm{u} = \frac{\mathrm{v}}{L} = (u_x, u_y, u_z)
\label{eq:segment_vector}
\end{equation}
Here, $\mathrm{u}$ is the local fiber axis direction used to compute misalignment at any slice crossed by the segment. In this work, the sampling resolution in the fiber direction is approximately $4~\mu\text{m}$.

Let $\mathrm{e}_z = (0,0,1)$ be the unit vector along the global $Z$-axis. The inclination of the local fiber axis relative to $Z$ is computed from the dot product, resulting in the depth misalignment with respect to the $Z$-axis, as given in Eq.~\eqref{eq:theta_z_ellipse}:
\begin{equation}
\theta_Z = \arccos\left( \left| \mathrm{u} \cdot \mathrm{e}_z \right| \right)
= \arccos\left( |u_z| \right)
\label{eq:theta_z_ellipse}
\end{equation}
where $\theta_Z \in [0^\circ,90^\circ]$. The absolute value ensures that the inclination is independent of whether the local segment points upward or downward along $Z$, which is appropriate when only the magnitude of tilt is required.

Signed in-plane misalignment components, $\theta_X$ and $\theta_Y$, are defined as the planar tilts in the $ZX$ and $ZY$ planes, respectively. These angles are measured from the positive $Z$ direction toward the positive $X$ and $Y$ directions and are computed directly from the \texttt{atan2} formulation for numerical robustness and correct quadrant identification, as follows:
\begin{equation}
\theta_X = \operatorname{atan2}(u_x, u_z),
\qquad
\theta_Y = \operatorname{atan2}(u_y, u_z)
\label{eq:theta_xy_ellipse}
\end{equation}

More details on the practical derivation are provided in the Supplementary Material (Section 1).

\subsubsection{Central difference method}

To validate the ellipse-intersection method, the tangent of each fiber segment is computed via central-difference increments between neighboring nodes, as shown in Fig.~\ref{fig2}(b):
\begin{equation}
\Delta X_i = X_{i+1} - X_{i-1}, \qquad
\Delta Y_i = Y_{i+1} - Y_{i-1}, \qquad
\Delta Z_i = Z_{i+1} - Z_{i-1}
\label{eq:cdm_deltas}
\end{equation}

Then, the local slopes in the global frame are estimated as
\begin{equation}
\frac{dX}{dZ} \approx \frac{\Delta X_i}{\Delta Z_i},
\qquad
\frac{dY}{dZ} \approx \frac{\Delta Y_i}{\Delta Z_i}
\label{eq:cdm_slopes}
\end{equation}
relating them to the misalignment through
\begin{equation}
\theta_X = \tan^{-1}\left(\frac{dX}{dZ}\right),
\qquad
\theta_Y = \tan^{-1}\left(\frac{dY}{dZ}\right)
\label{eq:cdm_theta_xy}
\end{equation}

For the depth-direction misalignment, at point $P_i$ of a segment, the unnormalized tangent expressed in global coordinates is $\mathrm{u}_i$, and the standard basis vector for the $Z$-axis is $\mathbf{e}_z$:
\begin{equation}
\mathrm{u}_i =
\begin{bmatrix}
dX/dZ \\
dY/dZ \\
1
\end{bmatrix},
\qquad
\mathbf{e}_z =
\begin{bmatrix}
0 \\
0 \\
1
\end{bmatrix}
\label{eq:cdm_vectors}
\end{equation}

Using the dot product between the two vectors gives
\begin{equation}
\cos(\theta_Z)
=
\frac{\mathrm{u}_i \cdot \mathrm{e}_z}{\lVert \mathrm{u}_i \rVert \lVert \mathrm{e}_z \rVert}
=
\frac{1}{\sqrt{1 + \left(dX/dZ\right)^2 + \left(dY/dZ\right)^2}}
\label{eq:cdm_cos}
\end{equation}

Using the trigonometric identity
\begin{equation}
r = \sqrt{\left( \frac{dX}{dZ} \right)^2 + \left( \frac{dY}{dZ} \right)^2}
\label{eq:r_def}
\end{equation}
then
\begin{equation}
\arccos\left(\frac{1}{\sqrt{1+r^2}}\right) = \arctan(r)
\label{eq:trig_identity}
\end{equation}
Thus,
\begin{equation}
\theta_Z
=
\tan^{-1}(r)
=
\tan^{-1}\left(
\sqrt{
\left( \frac{dX}{dZ} \right)^2
+
\left( \frac{dY}{dZ} \right)^2
}
\right)
\label{eq:theta_z_cdm}
\end{equation}

Proof 1 in the Supplementary Material (Section 1) explains the equivalence of the depth misalignment relation, $\theta_Z$, between the ellipse-intersection method and the central difference method.

Repeating the above steps for all fibers and all slice planes produces a per-slice set of misalignment distributions, $\theta_X$, $\theta_Y$, and $\theta_Z$. These values are stored together with the corresponding slice index and fiber identifier and are subsequently used to compute per-slice statistics, such as mean, variance, and tails/outliers, and to calibrate probabilistic models of misalignment for synthetic microstructure generation.

\subsection{Synthetic fiber misalignment modelling}
\label{sec:synthetic_modelling}
In this section, the processed misalignment dataset is leveraged to construct a stochastic generative model that produces synthetic fibers whose slice-wise and along-fiber-length statistics match those of the original microstructure. The procedure is implemented sequentially, in which per-slice angle histograms are used to reproduce the marginal distributions of misalignment.

First, the within-slice dependence between $\theta_X$ and $\theta_Y$ is captured using a Gaussian copula. Second, along-fiber-length memory is introduced through an autoregressive formulation, ensuring that synthetic fibers retain realistic path continuity in the fiber direction. Third, to mitigate the tail-independence behavior inherent to Gaussian dependence models, where extreme events tend to behave nearly independently despite strong central correlation, an additional tail treatment is incorporated. Fourth, rare fibers exhibiting exceptionally high misalignment (motifs) are identified in the experimental dataset and re-injected into the synthetic population using a probabilistic resampling strategy. Finally, a Bayesian optimization step is employed to tune key numerical coefficients associated with the autoregressive and tail components, yielding a calibrated model capable of generating statistically equivalent synthetic microstructures that closely mirror the original data.

\subsubsection{Bivariate Gaussian copula model}

\begin{figure}[H]
    \centering
    \includegraphics[width=\textwidth]{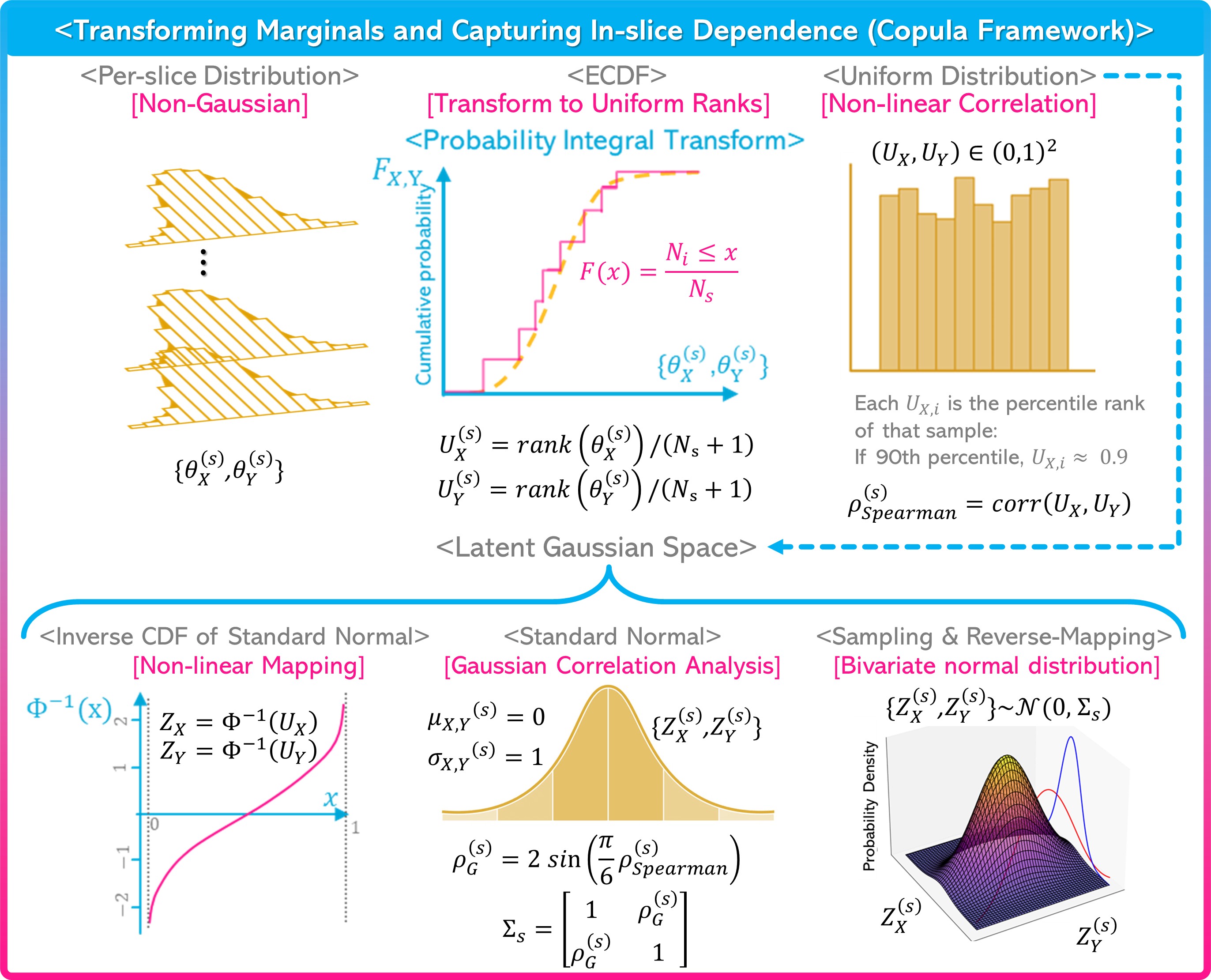}
    \caption{A schematic of the framework for marginals transformation and modelling the per-slice joint ($\theta_X$, $\theta_Y$) with Gaussian copula.}
    \label{fig3}
\end{figure}

Misalignment angles exhibit planar dependence on a per-slice basis; therefore, this dependence is modelled separately from the marginal distributions using a Gaussian copula. The effect of different copula types is beyond the scope of the present study. A schematic of the framework is illustrated in Fig.~\ref{fig3}.

The framework utilizes a standard probability integral transform, in which the marginal distributions are mapped to a uniform space using ranked data or the empirical cumulative distribution function (ECDF). The ECDF serves as a non-parametric step function that estimates the cumulative probability of a sample by calculating the proportion of observations less than or equal to a specific value. To ensure that transformed values remain strictly within the interval $(0,1)$, a rank-based formula, specifically the Weibull plotting position, is applied, converting sorted empirical data into estimates of their cumulative probability. The resulting uniform space preserves the ordinal structure of the data and facilitates the computation of the copula, ensuring that it remains invariant under monotonic transformations of the marginals.

On the $(U_X^{(s)}, U_Y^{(s)})$ pairs, a dependence measure is estimated using Spearman's $\rho$, which yields a value in $[-1,1]$ describing how tightly $U_X$ and $U_Y$ vary together. Then, the inverse cumulative distribution function (CDF) of the standard normal distribution is used as a nonlinear mapping to transform the per-slice marginals into the standard Gaussian space. This rank-based correlation is then mapped to a Gaussian-copula correlation using the standard monotonic relationship between Spearman correlation and Gaussian correlation, as follows \cite{ref40}:
\begin{equation}
\rho_G^{(s)} = 2 \sin\left( \frac{\pi}{6} \rho_{\mathrm{Spearman}}^{(s)} \right)
\label{eq:rho_gaussian}
\end{equation}

It is noted that the Gaussian copula definition follows:
\begin{equation}
C_{\rho}\left(\theta_X,\theta_Y\right)
=
\Phi_{\rho}\left(
\Phi^{-1}\left(\theta_X\right),
\Phi^{-1}\left(\theta_Y\right)
\right)
\label{eq:gaussian_copula}
\end{equation}
where $\Phi_{\rho}$ is the CDF of the joint standard bivariate normal distribution with correlation $\rho_G$, and $\Phi^{-1}$ is the inverse CDF of the standard univariate normal distribution.

Then, from the per-slice latent covariance matrix, the Cholesky lower-triangular decomposition is computed as follows:
\begin{equation}
\Sigma_{s} =
\begin{bmatrix}
1 & \rho_G^{(s)} \\
\rho_G^{(s)} & 1
\end{bmatrix}
\label{eq:sigma_slice}
\end{equation}

\begin{equation}
L_{s} = \operatorname{chol}\left(\Sigma_{s}\right)
=
\begin{bmatrix}
1 & 0 \\
\rho_G^{(s)} & \sqrt{1-\left(\rho_G^{(s)}\right)^2}
\end{bmatrix}
\label{eq:cholesky_slice}
\end{equation}

For a given fiber at a given slice $s$, a synthetic angle pair can be drawn by first sampling an independent standard normal vector as:
\begin{equation}
\varepsilon_{s} =
\begin{bmatrix}
\varepsilon_X \\
\varepsilon_Y
\end{bmatrix}
\sim \mathcal{N}(0, I_2)
\label{eq:epsilon_standard}
\end{equation}

and then correlating it with the Cholesky factor in the latent space as:
\begin{equation}
\mathrm{Z}_{s} =
\begin{bmatrix}
{\mathrm{Z}_X} \\
{\mathrm{Z}_Y}
\end{bmatrix}
=
L_{s} \varepsilon_{s}
\label{eq:zgen}
\end{equation}

This ensures that $(Z_X, Z_Y) \sim \mathcal{N}(0,\Sigma_{s})$, as shown in Proof 2 in the Supplementary Material (Section 2). In the case of mapping back from the latent Gaussian space to the original $\theta$-space, the variables are first mapped to the uniform space and then to angles using the inverse ECDF.

\subsubsection{Depth continuity via AR(1) model in latent space}

So far, each slice can produce the correct $(\theta_X,\theta_Y)$ independently, but real fibers vary smoothly along the fiber direction, with no sudden abrupt changes resembling a zig-zag motion. In other words, fibers exhibit a form of memory of their previous spatial positions. Thus, in this section, samples in the latent space are correlated along the fiber depth direction through a latent Gaussian-space correlation of the sampled $(Z_X,Z_Y)\sim\mathcal{N}(0,\Sigma_{s})$. This is achieved using a first-order autoregressive process, i.e., AR(1). In general, an autoregressive (AR) process is a model in which the next value depends on its own previous value(s), as follows \cite{ref41,ref42,ref43}:
\begin{equation}
Z_t = \phi_1 Z_{t-1} + \phi_2 Z_{t-2} + \cdots + \phi_p Z_{t-p} + \text{noise}
\label{eq:ar_general}
\end{equation}

In this work, for a fixed slice index $s$, a first-order autoregressive framework is defined as:
\begin{equation}
Z_{s}
=
\phi ~Z_{s-1}
+
\sqrt{1-\phi^2}\, Z_{\mathrm{noise}}^{(s)},
\quad
\text{where }
Z_{\mathrm{noise}}^{(s)} = L_{s} \varepsilon_{s},
\quad
\varepsilon_{s} \sim \mathcal{N}(0,I_2),
\quad
\phi \in [0,1)
\label{eq:ar1_latent}
\end{equation}

Here, $Z_{s} \in \mathbb{R}^2$ is the latent Gaussian variable for slice $s$, and $\phi$ is the AR(1) memory coefficient, which is described as 

\begin{equation}
\phi
=
\phi_{global}
+
\sigma_{\phi}\,\eta_i,
\qquad
\eta_i \sim \mathcal{N}(0,1),
\label{phi_global_local}
\end{equation}

\noindent where $\sigma_{\phi}$ is the \emph{jitter} standard deviation. This means that $\phi = \phi_{\mathrm{local},i}$, is  fiber-by-fiber local lag-1 coefficient used in the AR(1) recursion for each fibre \emph{i} rather than the fixed global value. This two-parameter representation $(\phi_{\mathrm{global}},\,\sigma_{\phi})$ is more expressive than a single shared coefficient: $\phi_{\mathrm{global}}$ controls the ensemble-mean smoothness of all fibres, while $\sigma_{\phi}$ controls the spread of individual fibre behaviors around that mean, producing a realistic population of fibres with varying degrees of curvature.

\begin{figure}[H]
    \centering
    \includegraphics[width=\textwidth]{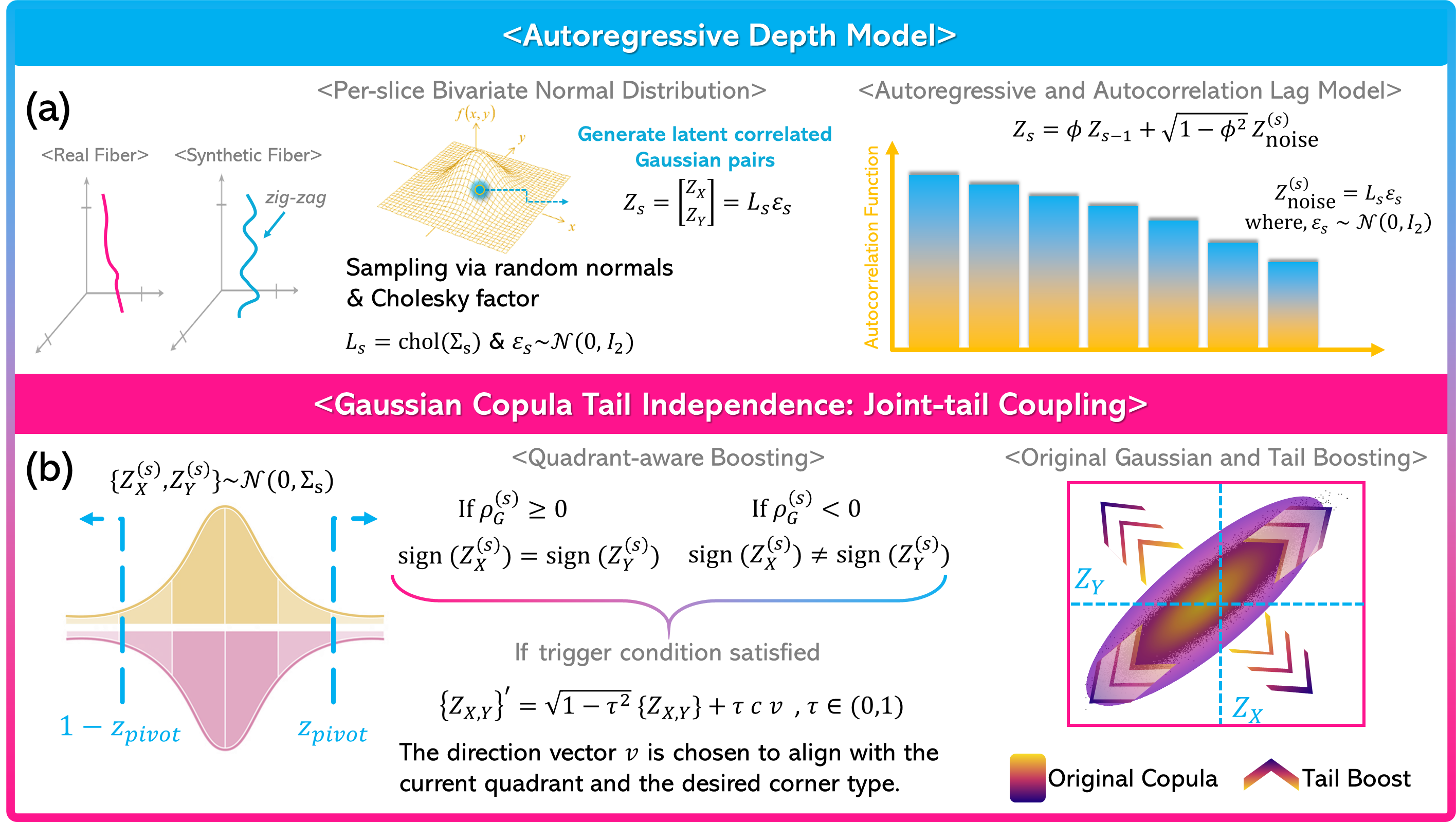}
    \caption{(a) A schematic of the autoregressive depth model. (b) Gaussian copula tail boosting for joint tail coupling cases.}
    \label{fig4}
\end{figure}

The term $L_{s}\varepsilon_{s}$ is the new correlated noise for this slice, while $\varepsilon_{s}$ is standard Gaussian white noise. The factor $\sqrt{1-\phi^2}$ is a variance-correction term, which is discussed together with the autocorrelation lag model in greater detail in Proof 3 of the Supplementary Material (Section 2). An illustration of the framework is shown in Fig.~\ref{fig4}(a).

In an AR(1) model, the autocorrelation at each lag represents the ``memory'' of the system, where a positive correlation indicates that the series exhibits persistence. This means that the data resist sudden changes; the Gaussian vector $(Z_X, Z_Y)$ of a fiber at slice $s$ inherits a fraction of the momentum from the previous slice, $s-1$. Consequently, the series evolves in smooth waves rather than through erratic jumps.

\subsubsection{Gaussian copula tail independence}
\label{Gaussian copula tail independence}
Gaussian copulas are attractive because they preserve prescribed marginal distributions and capture bulk dependence through a latent Gaussian correlation. However, they are tail independent, meaning that they tend to underestimate the co-occurrence of joint extremes, such as simultaneous per-slice large misalignments. To address this limitation, a tail-gated coupling is introduced in the latent Gaussian space that activates only when samples enter the extreme quantiles and injects a small, directionally consistent common perturbation, as shown in Fig.~\ref{fig4}(b). This selectively strengthens dependence in the joint tails while leaving the central, non-extreme behavior essentially unchanged, thereby providing a simple and tunable mechanism to better reproduce extreme co-misalignment without substantially altering the prescribed marginals. A more detailed explanation of the framework and its derivation is provided in the Supplementary Material (Section 2).

\subsubsection{Persistent high-misalignment runs ``motifs''}
\label{Persistent high-misalignment runs ``motifs''}
The AR(1) latent process promotes depth-wise smoothness but remains largely Gaussian in its extreme-event structure. Thus, in the previous section, we have introduced a gated local per-slice extreme co-misalignment mechanism. However, in measured fiber microstructures, high misalignment typically appears as structured, contiguous segments ``motifs'' rather than isolated outliers. These are considered as fibers spanning multiple slices with the same highly-misaligned behavior. To reproduce these non-Gaussian, spatially coherent events, motifs are detected in the original reference data and re-injected during synthesis using a structured container referred to as the motif library.

First, motifs are detected in the original microstructure. For each fiber and slice $s$, a robust slice-wise standardized exceedance score is computed based on the $\theta_Z$ marginal:
\begin{equation}
z_{\mathrm{score}}(i)
=
\frac{\left|\theta_Z(i)-p_{50}^{(s)}\right|}{\mathrm{IQR}_Z^{(s)}}
\label{eq:zscore_motif}
\end{equation}
where $p_{50}^{(s)}$ is the slice median of $\theta_Z$ and $\mathrm{IQR}_Z^{(s)}$ is the corresponding interquartile range. A motif candidate is declared when the following two conditions are simultaneously satisfied:
\begin{equation}
z_{\mathrm{score}}(i) > K~({\mathrm{IQR}})
\label{eq:motif_condition1}
\end{equation}
\begin{equation}
L_m^f = i_{\mathrm{end}} - i_{\mathrm{start}} + 1 \geq L_{\mathrm{threshold}}^f
\label{eq:motif_condition2}
\end{equation}

\begin{figure}[H]
    \centering
    \includegraphics[width=\textwidth]{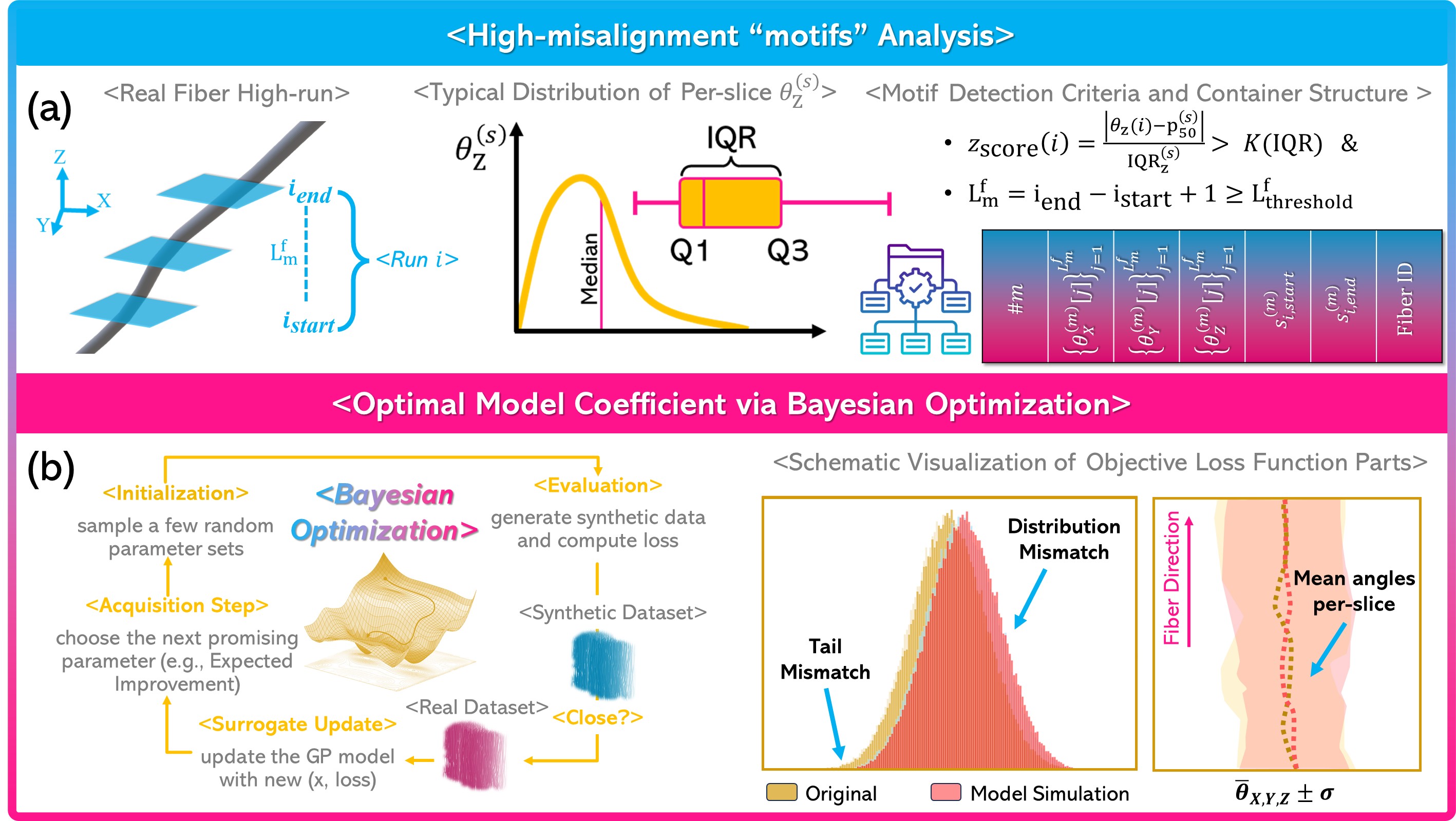}
    \caption{A schematic illustration of the (a) high-misalignment “motifs” analysis. (b) Bayesian optimization-based framework for model coefficients tuning.}
    \label{fig5}
\end{figure}

In this work, $K= 1$. This implies that the exceedance persists for at least $L_{\mathrm{threshold}}^f$ consecutive depth indices, where $L_m^f$ is the detected motif run for a single fiber, it is to be noted that, a single fiber can have muliple motif runs. $L_{\mathrm{threshold}}^f$ is set to the representative median motif length over all observed motifs in the scanned volume. Each detected run is stored as a motif $m$, together with the associated angle sequences and metadata, such as the motif planar misalignment, fiber-direction misalignment, $z_{\mathrm{score}}$, motif length, starting slice, ending slice, and percentile statistics. A schematic of this process is illustrated in Fig.~\ref{fig5}(a). It should be noted that motifs are detected fiber-by-fiber, rather than per slice, ensuring that contiguity is assessed along the fiber depth trajectory.

To characterize how frequently motifs occur as a function of slice index, the following quantities are computed: (i) an exceedance frequency,
\begin{equation}
p_{\mathrm{exceed}}^{(s)}
=
\frac{
\#\left\{
i \in \text{slice } s :
z_{\mathrm{score}}(i) > K~{\mathrm{(IQR)}}
\ \text{and} \
L_m^f \geq L_{\mathrm{threshold}}^f
\right\}
}{
N_s
}
\label{eq:pexceed}
\end{equation}
where $N_s$ is the number of available fiber samples at slice $s$. Its important to note this is referred as exceedance frequency as it is computed per slice counting how many fibers do actually fulfill the above two conditions, so a single motif run can be multiply counted in multiple slices although its the same motif. A (ii) a motif-start probability that avoids multiple-counting motifs spanning multiple slices, by characterizing every motif by its start slice index: 
\begin{equation}
\mathrm{Prob}_{\mathrm{motif\mbox{-}start}}^{(s)}
=
\frac{
\#\{\text{motifs that start at slice } s\}
}{
N_s
}
\label{eq:pmotif_start}
\end{equation}

To decide motif activation during fiber modelling and generation, motif injection is controlled by the slice-dependent probability $\mathrm{Prob}_{\mathrm{motif\mbox{-}start}}^{(s)}$. At slice $s$, a random variable $u \sim \mathcal{U}(0,1)$ is drawn, corresponding to a standard Monte Carlo decision process, and a motif is activated when
\[
u < \mathrm{Prob}_{\mathrm{motif\mbox{-}start}}^{(s)},
\]
provided that (i) the generator is not already inside an active motif and (ii) $\mathrm{Prob}_{\mathrm{motif\mbox{-}start}}^{(s)} > 0$. When activated, a motif instance is selected from a shuffled motif library to avoid repeated ordering artifacts. This way its stochastic in the sense that, for a certain slice the real and simulated microstructure do not have to own the same motif features, ensuring randomness and generalization of the model while still respecting the statistical behavior of the original data. 

To blend the motif contribution on top of the previously generated Gaussian angles, a smooth blending technique is applied. Let $\theta_{X,\mathrm{base}}$ and $\theta_{Y,\mathrm{base}}$ denote the baseline angles produced by the copula stage, AR(1), and conditionally tail-gated dependence enhancement, and let $\theta_{X,\mathrm{motif}}$ and $\theta_{Y,\mathrm{motif}}$ denote the angles from the selected motif. While a motif is active, the synthesized angles are blended as
\begin{equation}
\theta_X \leftarrow (1-\lambda)\theta_{X,\mathrm{base}} + \lambda \theta_{X,\mathrm{motif}}
\label{eq:blend_x}
\end{equation}
\begin{equation}
\theta_Y \leftarrow (1-\lambda)\theta_{Y,\mathrm{base}} + \lambda \theta_{Y,\mathrm{motif}}
\label{eq:blend_y}
\end{equation}

To avoid abrupt jumps at motif onset, $\lambda$ is ramped gradually over a fixed number of blending steps:
\begin{equation}
\lambda_i = \frac{i}{\mathrm{BlendSteps}},
\qquad
\text{where }
\mathrm{BlendSteps} = L_{\mathrm{threshold}}^f
\label{eq:lambda_blend}
\end{equation}
where $i$ is the local index since motif activation. This strategy injects rare, structured misalignment events while maintaining continuity and preserving the overall slice-level behavior outside the tails.

Finally, the generated angles are mapped back to spatial fiber coordinates. After generating $\theta_X(i)$ and $\theta_Y(i)$ on the same depth grid as the reference, with increments $\Delta Z$, the three-dimensional fiber centerline is reconstructed via
\begin{equation}
X_{i+1} = X_i + \tan\!\left(\theta_X(i)\right)\Delta Z
\label{eq:x_reconstruct}
\end{equation}
\begin{equation}
Y_{i+1} = Y_i + \tan\!\left(\theta_Y(i)\right)\Delta Z
\label{eq:y_reconstruct}
\end{equation}
with
\begin{equation}
Z_{i+1} = Z_i + \Delta Z
\end{equation}

This geometric step converts the synthesized local orientation field, expressed through the angles, into a spatial trajectory, thereby yielding a continuous three-dimensional fiber path consistent with the prescribed depth discretizations.

\subsubsection{Bayesian-optimization-based hyperparameter tuning}

The proposed generator contains a small set of hyperparameters that control (i) depth-wise persistence in the latent Gaussian process and (ii) the strength and activation of the joint-tail correction used to mitigate Gaussian tail independence. These parameters are fine-tuned via Bayesian optimization, namely $\phi_{global}$, $\phi_{\mathrm{Jitter}}$, $U_{\mathrm{Pivot}}$, and $\tau$.

Here, $\phi_{\mathrm{global}}$ governs the lag-1 dependence (depth memory persistance), and $\phi_{\mathrm{Jitter}}$ controls stochastic variability per-slice per-fiber, while $U_{\mathrm{Pivot}}$ (pivot quantile, which is synonyms to $z_{pivot}$, used here for easier interpretation of quantiles in uniform space) and $\tau$ (coupling strength) regulate the co-rare event tail-coupling mechanism.

To select these parameters systematically, a scalar discrepancy objective, $\mathcal{L}$, is minimized using Bayesian optimization (BO). At each BO iteration, candidate hyperparameters are proposed by a probabilistic surrogate model, i.e. a Gaussian process. The microstructure modelling pipeline is then executed using these parameters, and the evaluation metrics are computed by comparing the generated and reference microstructures. The surrogate model is subsequently updated with the resulting objective value. This procedure is repeated until a prescribed number of evaluations is reached (100 iterations in this work), yielding the hyperparameters that best match the target statistics.

At each iteration, a complete synthetic microstructure of fiber centrelines is generated, and the synthetic and original reference microstructures are compared using complementary metrics that capture marginal distributions, depth-wise trends, dependence, and extremes, as follows.

\paragraph{Kolmogorov-Smirnov (KS) distance}
The Kolmogorov--Smirnov (KS) distance is computed per angle component $\theta_X$, $\theta_Y$, and $\theta_Z$ to quantify marginal agreement:
\begin{equation}
D_{\mathrm{KS}}(\mathrm{orig},\mathrm{synth})
=
Max_{(~x= \theta_X,\theta_Y,\theta_Z)}
\left|
CDF_{\mathrm{orig}}(x)-CDF_{\mathrm{synth}}(x)
\right|
\label{eq:ks_distance}
\end{equation}
This metric captures the largest vertical gap differences in the overall shape of the marginal distributions between the real and synthetic data based on the CDF curves. It is computed globally for the full microstructure for each angle component, and is sensitive to differences in skewness and central mass even when the means and variances match.

\paragraph{Tail quantile error}
The tail quantile error directly penalizes mismatches in extreme misalignment magnitudes:
\begin{equation}
\mathrm{TailErr}
=
\frac{1}{3}
\sum_{\alpha \in \{X,Y,Z\}}
\frac{
\dfrac{1}{|Q|}
\displaystyle\sum_{q\,\in\,Q}
\left|
Q_q\!\left(\theta_{\alpha}^{\mathrm{synth}}\right)
-
Q_q\!\left(\theta_{\alpha}^{\mathrm{orig}}\right)
\right|
}{
\operatorname{range}\!\left(\theta_{\alpha}^{\mathrm{orig}}\right)
}
\label{eq:tail_error}
\end{equation}

where $Q = \{0.01,\,0.05,\,0.10,\,0.90,\,0.95,\,0.99\}$  This metric captures mismatches in the high-misalignment tails of the synthetic angle distributions relative to the original, evaluated symmetrically across both the upper ($q \in \{0.90, 0.95, 0.99\}$) and lower ($q \in \{0.01, 0.05, 0.10\}$) 
quantile levels for each of the three angle components $\theta_X$, $\theta_Y$, and $\theta_Z$. It is included as a dedicated term in the BO objective because global goodness-of-fit statistics such as the KS distance are most sensitive at the median of the distribution and can tolerate substantial errors in the extreme quantiles while still reporting a low loss. Since the primary motivation for the joint-tail coupling stage is to correct the systematic under-representation of large misalignment angles that arises from the tail-independence of the Gaussian copula, a tail-specific penalty is necessary to steer the BO towards hyperparameter combinations that genuinely recover the observed frequency of extreme fibre angles. Each per-component error is normalised by the empirical range of the original distribution to make the penalty dimensionless and comparable across components with different angle scales, and the final metric is the mean over all three components and all six quantile levels.

\paragraph{Depth-profile normalised root-mean-square error}
The depth-profile normalised root-mean-square error (NRMSE) penalises mismatches between the per-slice mean angle profiles of the synthetic and original microstructures:

\begin{equation}
\mathrm{NRMSE}_{\alpha}
=
\frac{
\sqrt{
\dfrac{1}{S}
\displaystyle\sum_{s=1}^{S}
\left(
\bar{\theta}_{\alpha,\mathrm{synth}}^{(s)}
-
\bar{\theta}_{\alpha,\mathrm{orig}}^{(s)}
\right)^2
}
}{
\operatorname{range}\!\left(
\bar{\theta}_{\alpha,\mathrm{orig}}
\right)
},
\qquad
\alpha \in \{X,Y,Z\}
\label{eq:nrmse_depth}
\end{equation}

\noindent where $\bar{\theta}_{\alpha}^{(s)}$ denotes the mean angle of component $\alpha$ across all fibres in slice $s$, $S$ is the number of slices, and the denominator is the range of the original mean profile $\{\bar{\theta}_{\alpha,\mathrm{orig}}^{(s)}\}_{s=1}^{S}$. The scalar loss contribution entering the BO objective is the worst-case component as $\max_{\alpha \in \{X,Y,Z\}}
\mathrm{NRMSE}_{\alpha}$. A depth-wise trend failure in any single component is fully penalized regardless of how well the other two components perform.

\paragraph{Per-slice copula correlation deviation}
To match slice-wise bivariate dependence structure, the per-slice Gaussian copula correlation copula deviation is then defined as the mean absolute difference between the synthetic and original Gaussian copula correlations over all slices:
\begin{equation}
{\left|\Delta \rho\right|}
=
\frac{1}{S}
\sum_{s \in \mathcal{S}}
\left|
\rho_{\mathrm{synth}}^{(s)}
-
\rho_{\mathrm{orig}}^{(s)}
\right|
\label{eq:copula_deviation}
\end{equation}

\noindent This metric captures the mismatch in $XY$ bivariate dependence structure between the synthetic and original microstructures independently of the marginal distributions; it may be small even when the marginals do not closely match, or vice versa.

\paragraph{Joint tail mismatch}
The joint tail mismatch metric directly penalizes errors in the co-occurrence of simultaneously extreme misalignment angles, i.e.\ the tendency of $\theta_X$ and $\theta_Y$ to be extreme together. Four corner quadrant probability masses are then estimated for a threshold $q$:
\begin{equation}
\hat{p}_{++}(q) = P(U_X > q,\; U_Y > q), \quad
\hat{p}_{--}(q) = P(U_X < 1{-}q,\; U_Y < 1{-}q),
\label{eq:same_tail_quads}
\end{equation}
\begin{equation}
\hat{p}_{+-}(q) = P(U_X > q,\; U_Y < 1{-}q), \quad
\hat{p}_{-+}(q) = P(U_X < 1{-}q,\; U_Y > q).
\label{eq:opp_tail_quads}
\end{equation}
The first pair ($\hat{p}_{++}$, $\hat{p}_{--}$) captures concordant co-occurrences;  both components simultaneously in their upper or lower extreme tails; while the second pair ($\hat{p}_{+-}$, $\hat{p}_{-+}$) captures discordant co-occurrences. The metric is then the weighted sum of absolute differences between the synthetic and original quadrant masses,
\begin{equation}
\mathcal{L}_{\mathrm{jointtail}}
=
\left|\hat{p}_{++}^{\mathrm{synth}} - \hat{p}_{++}^{\mathrm{orig}}\right|
+
\left|\hat{p}_{--}^{\mathrm{synth}} - \hat{p}_{--}^{\mathrm{orig}}\right|
+
\left|\hat{p}_{+-}^{\mathrm{synth}} - \hat{p}_{+-}^{\mathrm{orig}}\right|
+
\left|\hat{p}_{-+}^{\mathrm{synth}} - \hat{p}_{-+}^{\mathrm{orig}}\right|
\label{eq:tail_copula}
\end{equation}

\noindent with $q = 0.90$, treating all four corner types equally since concordant and discordant extreme co-occurrences are physically equivalent in terms of total fiber direction misalignment magnitude.

These terms are combined into a single weighted objective:
\begin{equation}
\mathcal{L}
=
w_{\mathrm{KS}}\,D_{\mathrm{KS}}
+
w_{\mathrm{NRMSE}}\,\mathrm{NRMSE}
+
w_{\mathrm{TE}}\,\mathrm{TailErr}
+
w_{\rho}\,{\left|\Delta\rho\right|}
+
w_{\mathrm{JT}}\,\mathrm{JointTail}
\label{eq:bo_objective}
\end{equation}

This scalar loss enables automated calibration by simultaneously enforcing agreement in (i) marginal distributions, (ii) depth-wise behavior, (iii) slice-level dependence, and (iv) joint-extreme statistical properties, all of which are required for faithful synthetic microstructure reproduction. In the current work, all the coefficients are equal, i.e. all metrics have similar significance.

\subsection{Physical microstructure generation}
\label{sec:synthetic_generation}

So far, statistical modelling has been employed to construct a data-informed generative layer that reproduces statistically equivalent characteristics of the measured microstructure. While the statistical layer provides realistic misalignment fields, the present section focuses on the geometric packing and overlap-resolution procedure that converts those statistics into physically admissible fiber arrangements. In this section, the statistical layer is embedded into a physical-space synthesis workflow. First, an initial seeding layer is populated with circular fiber cross-sections whose radii are sampled to reflect the observed variability of a real CFRP composite. Second, a slice-by-slice three-dimensional growth procedure is adopted in which, at each depth increment, an online version of the model developed in Section~\ref{sec:synthetic_modelling} is used to propose fiber-direction candidates and corresponding misalignments. Physical feasibility is enforced at every slice by guaranteeing a no-overlap configuration through an iterative ellipse non-overlap solver based on a projected Gauss--Seidel (PGS) scheme.

\subsubsection{Stage I: Initial 2D packing seeding layer }

The slice-by-slice three-dimensional growth algorithm requires an initial, physically admissible cross-sectional configuration at the base plane. A seed packing is therefore generated at $z=0$, in which each fiber is represented by a circular disk with radius $R_i$ and center
\[
\mathrm{c}_i(0)=
\begin{bmatrix}
x_i(0)\\
y_i(0)
\end{bmatrix}
\]
The radii are sampled from the prescribed radius distribution observed in one of authors' previous work \cite{ref44,ref45} in order to reproduce realistic polydispersity. The goal of Stage I is to construct a set of centers $\left\{\mathbf{c}_i(0)\right\}$ that (i) satisfies the target number of requested fibers and (ii) enforces strict non-overlap between all circular cross-sections. At this initialization step, fibers are assumed to be vertical, such that the in-plane inclination angles are set to $\theta_{X,i}(0)=\theta_{Y,i}(0)=0$, and hence the initial cross-sections are circular. The resulting $z=0$ configuration serves as the geometric seed for the subsequent three-dimensional trajectory reconstruction.

To impose the non-overlap constraint, for any pair of fibers $(i,j)$, the inter-center distance is defined as
\begin{equation}
d_{ij}=\lVert \mathrm{c}_i-\mathrm{c}_j \rVert_2
\label{eq:dij_seed}
\end{equation}
A feasible packing must satisfy the pairwise separation constraint
\begin{equation}
d_{ij}\geq R_i+R_j+\delta
\label{eq:seed_nonoverlap}
\end{equation}
where $\delta \geq 0$ is an optional clearance margin introduced to prevent near-contact numerical issues and to provide a minimal matrix gap if desired. In this work, $\delta$ is set to $0.3~\mu\text{m}$. Violations are quantified by the signed overlap measure
\begin{equation}
\Delta_{ij}=R_i+R_j+\delta-d_{ij}
\label{eq:seed_overlap}
\end{equation}
with $\Delta_{ij}\leq 0$ indicating feasibility and $\Delta_{ij}>0$ indicating overlap.

\paragraph{Phase I: Dynamic growth (DG)}

To obtain a dense yet disordered packing without excessive rejection, a dynamic growth strategy is first employed. In DG, fibers are initialized with small radii $r_i^{(0)} \ll R_i$ at provisional center locations $\mathbf{c}_i^{(0)}$. The radii are then increased gradually toward their target values while continuously resolving collisions. At DG iteration $k$, the growth step is
\begin{equation}
r_i^{k+\frac{1}{2}}=\min\left(R_i,\; r_i^k+\Delta r\right)
\label{eq:dg_growth}
\end{equation}
where $\Delta r$ controls the growth rate and is chosen to balance convergence speed and collision stability.

The notation $r_i^{k+\frac{1}{2}}$ follows a split-step, or operator-splitting, convention. It denotes an intermediate sub-step within one full DG iteration $k \rightarrow k+1$. The radii are advanced first, whereas the centers $\mathbf{c}_i^k$ are not yet updated, so the state is only partially advanced. It is therefore a bookkeeping label used to distinguish a state that is neither the start of iteration $k$ nor the fully resolved end-state $k+1$, but rather an intermediate state in which the radii have been advanced while the centers have not yet been corrected.

After the radii are updated at each step, overlaps are removed through an iterative collision-resolution loop. For each overlapping pair $(i,j)$ with $\Delta_{ij}>0$, the unit separation direction is computed as
\begin{equation}
\mathrm{n}_{ij}=
\frac{\mathrm{c}_i-\mathrm{c}_j}{\lVert \mathrm{c}_i-\mathrm{c}_j \rVert_2}
\label{eq:nij_dg}
\end{equation}
and a symmetric repulsion update is applied:
\begin{equation}
\mathrm{c}_i \leftarrow \mathrm{c}_i+\frac{1}{2}\alpha \Delta_{ij}\mathrm{n}_{ij},
\qquad
\mathrm{c}_j \leftarrow \mathrm{c}_j-\frac{1}{2}\alpha \Delta_{ij}\mathrm{n}_{ij}
\label{eq:repulsion_dg}
\end{equation}
where $0<\alpha\leq1$ is a damping factor that stabilizes the iterative corrections and mitigates oscillations in dense configurations. The correction loop is repeated, cycling through all relevant pairs, until all overlaps are removed for the current radii, i.e., $\Delta_{ij}\leq0$ for all pairs. The DG phase terminates when (i) $r_i=R_i$ for all fibers and (ii) the non-overlap constraint remains satisfied after collision resolution at the final radii. DG is particularly effective in polydisperse systems because gradual expansion allows the packing to self-organize into a dense configuration without imposing crystalline order, while the repulsion step provides a computationally simple mechanism to restore feasibility after each growth increment. In this work, a threshold of 90\% of the target fiber volume fraction or target number of fibers is used for the DG phase in order to balance computational time, achieve high packing density, and retain additional randomness for Phase II.

\paragraph{Phase II: Random sequential adsorption (RSA)}

\begin{figure}[H]
    \centering
    \includegraphics[width=\textwidth]{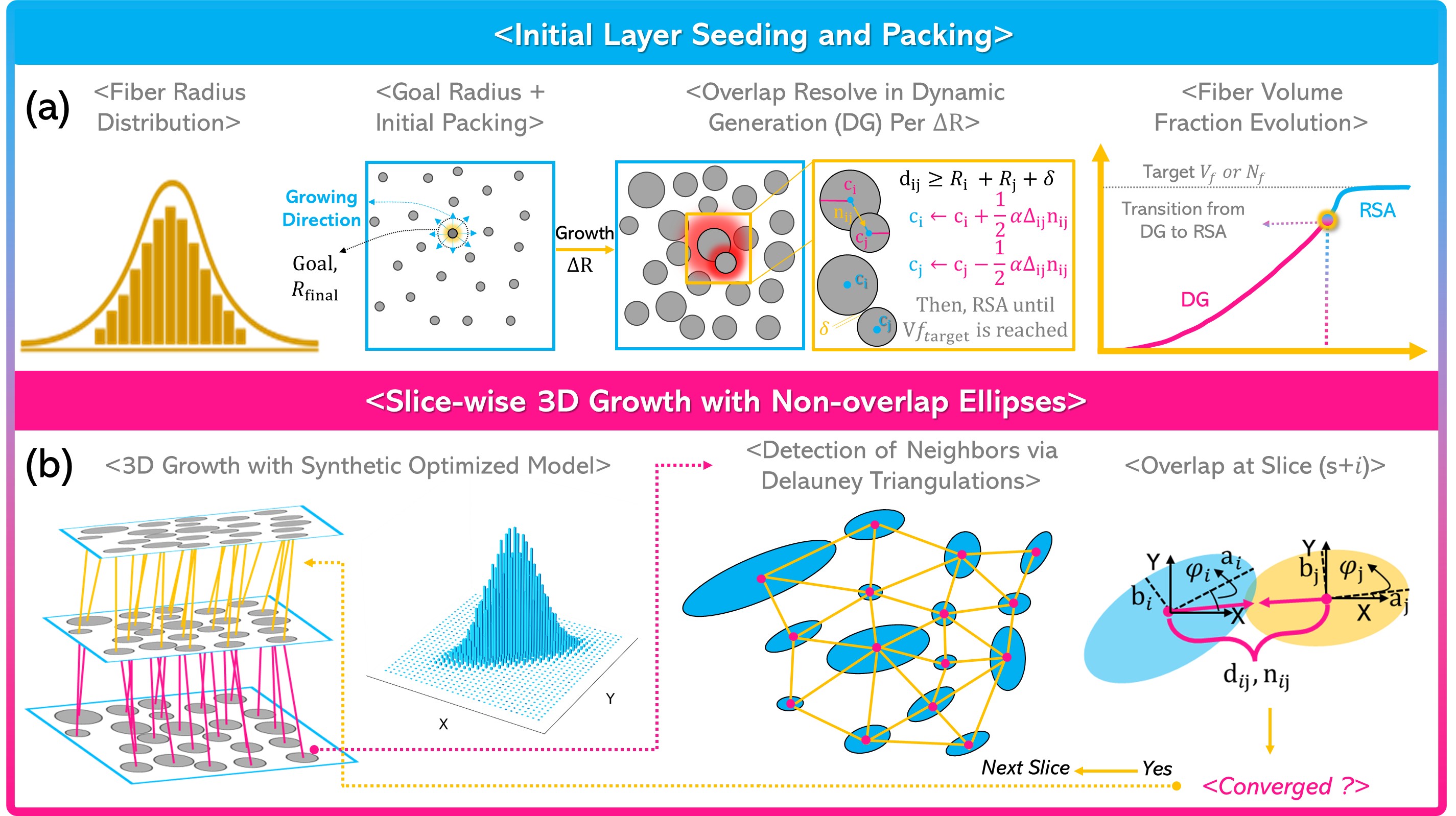}
    \caption{A schematic of the microstructure generation framework. (a) Initial layer microstructure generation. (b) slice-wise 3D growth and overlap detection and resolution for eclipses.}
    \label{fig6}
\end{figure}

After DG, residual void space may remain, especially in finite domains or when the target counts or volume fraction are not fully met due to growth-limited placement. A second-stage RSA procedure is therefore applied to populate the remaining space while preserving disorder. In RSA, a candidate fiber with radius $R_{\mathrm{cand}}$ is proposed by sampling a candidate center $\mathbf{c}_{\mathrm{cand}}$ uniformly within the domain. The candidate is accepted only if it satisfies
\begin{equation}
\lVert \mathbf{c}_{\mathrm{cand}}-\mathbf{c}_j \rVert_2
\geq
R_{\mathrm{cand}}+R_j+\delta,
\qquad \forall j
\label{eq:rsa_acceptance}
\end{equation}
otherwise, it is rejected and a new candidate location is sampled. RSA proceeds until the required number of fibers, or target areal fraction, is reached. A schematic of the framework is illustrated in Fig.~\ref{fig6}(a).

\subsubsection{Stage II: Slice-wise 3D growth with online statistical sampling and geometric feasibility enforcement}

Following the collision-free initialization at $z=0$, the fiber system is propagated through the depth using a slice-wise three-dimensional growth scheme. At each increment $s \rightarrow s+1$, with slice thickness $\Delta z$, the algorithm couples an online statistical generator calibrated in Section~\ref{sec:synthetic_modelling} with an ellipse-aware geometric solver to ensure that the evolving configuration remains physically admissible. Each slice is accepted and committed only after satisfying a strict no-overlap criterion; otherwise, the slice is resampled and the procedure is repeated. For a given slice transition $s \rightarrow s+1$, the workflow consists of: (i) sampling fiber-direction candidates from the calibrated statistical model and then selecting the most feasible candidate for each fiber at that slice, (ii) projecting centers to provisional positions at slice $s+1$, (iii) resolving inter-fiber overlaps iteratively using PGS contact solver with elliptical cross-sections, and (iv) performing a verification step prior to committing the slice.

\paragraph{Online statistical sampling}

Each fiber $i$ carries a latent bivariate Gaussian state, which evolves in depth according to a first-order autoregressive update:
\begin{equation}
\mathrm{Z}_i^{(s)} =
\begin{bmatrix}
Z_{i,x}^{(s)} \\
Z_{i,y}^{(s)}
\end{bmatrix},
\qquad
\mathrm{Z}_i^{(s+1)}
=
\phi_{local, i} \,\mathrm{Z}_i^{(s)}
+
\sqrt{1-\phi_{local,i}^2}\,
\left(L_{s+1} ~\varepsilon_{s+1}\right)
\label{eq:online_ar_update}
\end{equation}
where, optionally, the boosted joint-tail treatment described in Section~\ref{Gaussian copula tail independence} is applied.

The latent state is then mapped component-wise to the uniform domain using the standard normal cumulative distribution function:
\begin{equation}
U_{i,x}^{(s+1)}=\Phi\!\left(Z_{i,x}^{(s+1)}\right),
\qquad
U_{i,y}^{(s+1)}=\Phi\!\left(Z_{i,y}^{(s+1)}\right)
\label{eq:online_uniform_map}
\end{equation}

Finally, slice-wise empirical inverse CDFs are used to map the uniform variables to physical angles:
\begin{equation}
\theta_{X,i}^{(s+1)}
=
F_{\theta_X,s}^{-1}\!\left(U_{i,x}^{(s+1)}\right),
\qquad
\theta_{Y,i}^{(s+1)}
=
F_{\theta_Y,s}^{-1}\!\left(U_{i,y}^{ (s+1)}\right)
\label{eq:online_angle_map}
\end{equation}
Optionally, the generated angles are then blended with motifs as described in Section~\ref{Persistent high-misalignment runs ``motifs''}. These angles represent statistically consistent candidate orientations at the target depth and are generated online, or on-the-fly, at each slice using the calibrated per-slice marginals, dependence structure, statistical descriptors, and tuned hyperparameters.

\paragraph{Provisional center projection (pre-solver motion)}

Given the sampled angles $\theta_{X,i}^{(s+1)}$ and $\theta_{Y,i}^{(s+1)}$, the fiber center is advanced by a kinematic projection over the slice thickness $\Delta Z_{slice}$:
\begin{equation}
\Delta X_i^{(s)}=\tan\!\left(\theta_{X,i}^{(s+1)}\right)\Delta Z_{slice},
\qquad
\Delta Y_i^{(s)}=\tan\!\left(\theta_{Y,i}^{(s+1)}\right)\Delta Z_{slice}
\label{eq:delta_xy_projection}
\end{equation}
\begin{equation}
\mathrm{c}_i^{\mathrm{pre,~(s+1)}}
=
\mathrm{c}_i^{(s)}
+
\begin{bmatrix}
\Delta X_i^{(s)} \\
\Delta Y_i^{(s)}
\end{bmatrix}
\label{eq:center_preprojection}
\end{equation}

This produces a provisional configuration at slice $s+1$. If no geometric conflicts are detected, the slice can be accepted immediately; otherwise, an overlap-resolution stage is invoked, as described in the following section.

Once the final centers $\mathrm{c}_i^{(s)}$ and $\mathrm{c}_i^{(s+1)}$ are determined, the realized geometric angles are computed from the actual chord connecting consecutive slice positions:
\begin{equation}
\theta_{X,i}^{\mathrm{geom, (s)}}
=
\arctan\!\left(
\frac{x_i^{(s+1)}-x_i^{(s)}}{\Delta Z_{slice}}
\right),
\qquad
\theta_{Y,i}^{\mathrm{geom, (s)}}
=
\arctan\!\left(
\frac{y_i^{(s+1)}-y_i^{(s)}}{\Delta Z_{slice}}
\right)
\label{eq:geom_angles_realized}
\end{equation}

These realized angles constitute the physically meaningful fiber orientations and are used for ellipse construction, overlap detection, and final statistical validation.

\paragraph{Overlap resolution using an ellipse-aware PGS solver}

To enforce geometric feasibility at slice $s+1$, a constrained separation problem is solved in which ellipse pairs must satisfy a directional non-overlap condition. For a tilted cylindrical fiber intersected by a planar slice, the resulting cross-section is an ellipse. Accordingly, each fiber cross-section at slice $s+1$ is represented by an ellipse with semi-minor and semi-major axes
\begin{equation}
b_i = R_i,
\qquad
a_i
=
\left(
\frac{R_i}{|u_{z,i}|}
\right)
\label{eq:ellipse_axes}
\end{equation}
where the unit direction vector is formed from the realized tilts in Eq.~\eqref{eq:geom_angles_realized}, and overlap detection and resolution are applied only to the overlapping pairs in the pre-committed slice-center set $E$.

The unit direction vector between slices is given by
\begin{equation}
\mathrm{u}_i
=
\frac{1}{
\sqrt{
1+\tan^2\!\left(\theta_{X,i}^{\mathrm{geom}}\right)
+\tan^2\!\left(\theta_{Y,i}^{\mathrm{geom}}\right)
}
}
\begin{bmatrix}
\tan\!\left(\theta_{X,i}^{\mathrm{geom}}\right) \\
\tan\!\left(\theta_{Y,i}^{\mathrm{geom}}\right) \\
1
\end{bmatrix},
\qquad
u_{z,i}
=
\frac{1}{
\sqrt{
1+\tan^2\!\left(\theta_{X,i}^{\mathrm{geom}}\right)
+\tan^2\!\left(\theta_{Y,i}^{\mathrm{geom}}\right)
}
}
\label{eq:unit_direction_geom}
\end{equation}

The in-plane rotation of the ellipse is
\begin{equation}
\varphi_i
=
\operatorname{atan2}
\left(
\tan\!\left(\theta_{Y,i}^{\mathrm{geom}}\right),
\tan\!\left(\theta_{X,i}^{\mathrm{geom}}\right)
\right)
\label{eq:ellipse_rotation}
\end{equation}

The directional support radius along a unit vector $\mathrm{n}=(n_x,n_y)$ is computed by rotating $\mathrm{n}$ into the ellipse frame via $\mathrm{n}'=\mathrm{T}(\varphi)\mathrm{n}$, where $\mathrm{T}$ is the in-plane transformation matrix. The corresponding support function in the rotated local coordinate frame is
\begin{equation}
\rho(a,b,\varphi;\mathrm{n})
=
\frac{1}{
\sqrt{
\frac{(n_x')^2}{a^2}
+
\frac{(n_y')^2}{b^2}
}
}
\label{eq:support_radius}
\end{equation}
which gives the distance from the ellipse center to its boundary along direction $\mathrm{n}$.

For a candidate pair $(i,j)$, with inter-center distance
\[
d_{ij}=\lVert \mathrm{c}_i-\mathrm{c}_j \rVert_2
\]
and direction
\[
\mathrm{n}_{ij}=\frac{\mathrm{c}_i-\mathrm{c}_j}{d_{ij}},
\]
the required separation is expressed as
\begin{equation}
d_{ij}\geq \gamma~(r_i+r_j),
\qquad
r_i=\rho(a_i,b_i,\varphi_i;\mathrm{n}_{ij}),
\qquad
r_j=\rho(a_j,b_j,\varphi_j;-\mathrm{n}_{ij})
\label{eq:ellipse_separation}
\end{equation}
where $\gamma\geq1$ is a small safety factor, e.g.\ $\gamma=1.02$, introduced to guard against numerical tolerance and discretizations effects. The signed gap, or constraint violation, is defined as
\begin{equation}
g_{ij}
=
\gamma~(r_i+r_j)-d_{ij}
\label{eq:signed_gap}
\end{equation}
with $g_{ij}\leq0$ indicating feasibility and $g_{ij}>0$ indicating overlap.

The PGS solver iterates over a sparse set of neighboring pairs and applies symmetric corrections for any violated constraint. For each violating pair, the displacement increment is
\begin{equation}
\Delta
=
\frac{1}{2}\,\omega\,g_{ij}\,\mathrm{n}_{ij}
\label{eq:pgs_increment}
\end{equation}
where $\omega\in(0,1)$ is a relaxation parameter controlling update aggressiveness. The centers are then updated symmetrically:
\begin{equation}
\mathrm{c}_i \leftarrow \mathrm{c}_i+\Delta,
\qquad
\mathrm{c}_j \leftarrow \mathrm{c}_j-\Delta
\label{eq:pgs_center_update}
\end{equation}

A schematic of overlap detection and resolution for a neighboring pair is shown in Fig.~\ref{fig6}(b). 

PGS iterates for a fixed number of passes, or until the maximum violation becomes sufficiently small:
\begin{equation}
g_{ij,\max}<\varepsilon_{\mathrm{gap}}
\label{eq:pgs_convergence}
\end{equation}
A more detailed explanation of the implementation and verification frameworks are provided in the Supplementary Material (Section 3).

A key coupling exists between the centers and the ellipse geometry: the ellipse parameters $a_i$, $b_i$, and $\varphi_i$ depend on the realized geometric angles, which themselves depend on the updated centers. To handle this dependency, a fixed-point iteration is performed at each sampling round:
\begin{enumerate}
    \item compute the realized geometric angles from the current chord using Eq.~\eqref{eq:geom_angles_realized};
    \item build the ellipses, via $a_i$, $b_i$, and $\varphi_i$, from $\theta^{\mathrm{geom}}$;
    \item rebuild the neighbor set $E_{\mathrm{Del}}$ using Delaunay triangulation on the current centers $\mathbf{c}_i^{\mathrm{curr~(s+1)}}$;
    \item run a PGS solve with the geometry fixed within that pass, update $\mathbf{c}_i^{\mathrm{curr}}(s+1)$, and stop early if the maximum center displacement falls below a negligible threshold $\varepsilon_{\mathrm{pgs}}$.
\end{enumerate}

This inner fixed-point loop accounts for the fact that moving centers alters the chord direction, and hence the ellipse shape and orientation, which in turn changes the contact constraints. Convergence is reached when the centers and geometry are mutually consistent. Crucially, this loop does not constitute resampling: no statistical state is discarded or modified. If, after this loop, the subsequent all-pairs geometric verification still detects violations exceeding the tolerance $\varepsilon_{\mathrm{gap}}$, the entire round, including the proposed centers and associated latent state, is rejected without commitment, and a new statistical proposal is drawn. This outer resampling loop is the mechanism by which the algorithm enforces strict geometric feasibility at each slice.

After convergence, a strict verification step (Supplementary Material (Section 3)) evaluates geometric feasibility based on the final realized geometry, i.e., ellipses reconstructed using $\theta^{\mathrm{geom}}$. The slice is accepted only if all checked pairs satisfy the non-overlap condition. If verification fails, the slice is resampled, meaning that new statistical angles are drawn, and the projection$\rightarrow$solve$\rightarrow$verify cycle is repeated.

Once slice $s+1$ is accepted, the final centers $\mathbf{c}_i^{(s+1)}$ are committed, the latent AR states and motif states are advanced, the realized geometric angles are stored, and the algorithm proceeds to slice $s+2$. This procedure enforces, by construction: (i) geometric non-overlap at every depth, (ii) statistical consistency with the calibrated modelling layer, and (iii) physically meaningful fiber orientations derived from the realized centerline geometry.

\section{Results and discussion}

\subsection{Per-slice misalignment analysis of the original microstructure}

\begin{figure}[H]
    \centering
    \includegraphics[width=\textwidth]{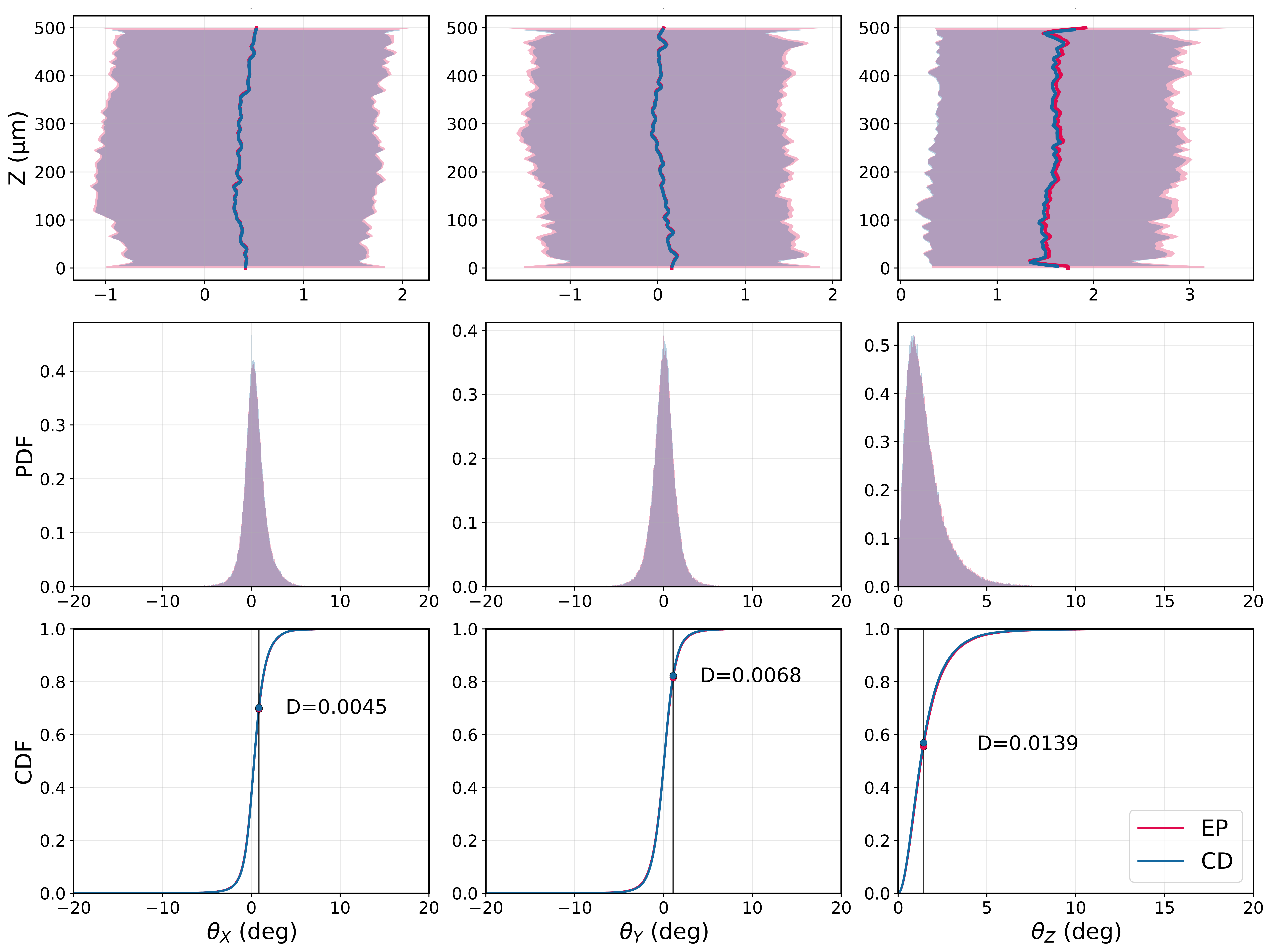}
    \caption {(a) Overlay of the aggregated mean $\pm \sigma$ versus depth for the ellipse-intersection method (EP, pink) and the central-difference method (CD, blue) for $\theta_X$, $\theta_Y$, and $\theta_Z$. (b) Overlaid aggregated misalignment distributions (PDFs) for $\theta_X$, $\theta_Y$, and $\theta_Z$ obtained using the two methods. (c) Empirical CDF comparison for $\theta_X$, $\theta_Y$, and $\theta_Z$, highlighting the maximum Kolmogorov--Smirnov (KS) separation between EP and CD.}
    \label{fig7}
\end{figure}

In this section, per-slice fiber misalignment is analyzed using the two methods described in Section~2.2. We first present the aggregated planar misalignment distributions, $\theta_X$ and $\theta_Y$, together with the through-thickness fiber-direction misalignment, $\theta_Z$, across all slices within the sub-volume. Fig.~\ref{fig7}(a) shows the depth-wise evolution of the aggregated mean $\pm \sigma$ for both methods, while Fig.~\ref{fig7}(b) overlays the corresponding aggregated probability density functions (PDFs). Fig.~\ref{fig7}(c) further compares the empirical cumulative distribution functions (CDFs) and indicates the location of the maximum Kolmogorov--Smirnov (KS) separation between the two methods.

The results demonstrate strong agreement between the ellipse-intersection and central-difference approaches in both the depth-wise statistics and the aggregated distributions of $\theta_X$, $\theta_Y$, and $\theta_Z$. The aggregated mean misalignments are nearly identical for $\theta_X$ (Ellipse: $0.387^\circ$ vs.\ CD: $0.386^\circ$) and $\theta_Y$ (Ellipse: $0.0499^\circ$ vs.\ CD: $0.0485^\circ$), whereas $\theta_Z$ exhibits a slightly larger but still modest offset (Ellipse: $1.601^\circ$ vs.\ CD: $1.563^\circ$). The CDF-based comparison corroborates this resemblance: the two-sample KS distances are small ($D = 0.00449$ for $\theta_X$, $D = 0.00684$ for $\theta_Y$, and $D = 0.0139$ for $\theta_Z$), corresponding to a maximum CDF separation of approximately $1.4\%$.

Minor differences are expected because the two methods are not strictly equivalent. The ellipse-based approach introduces approximations through ellipse parameterization and the use of a fixed monotonic radius $R$ when extracting misalignment, while the central-difference approach relies on a per-segment cylinder model that assumes locally straight fiber segments. These modelling choices primarily influence higher-order geometric details and curvature sensitivity, which plausibly contributes to the slightly larger deviation observed for $\theta_Z$. For the subsequent modelling and generation phases, the ellipse-intersection results are adopted because they explicitly incorporate $R$, providing a direct geometric descriptor of fiber shape.

\subsection{Synthetic microstructure fiber misalignment modelling}

\begin{figure}[H]
    \centering
    \includegraphics[width=\textwidth]{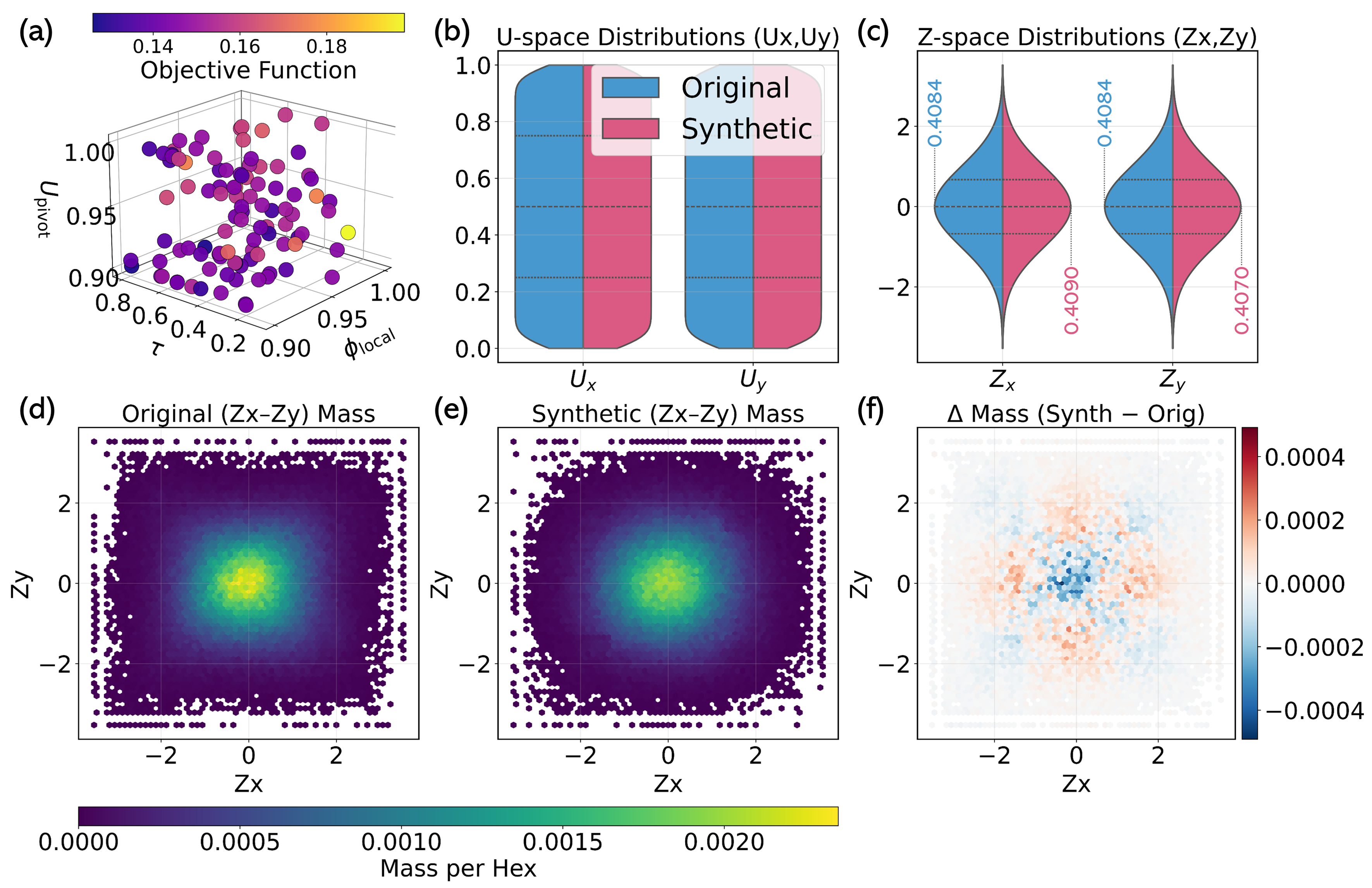}
    \caption{Bayesian optimization (BO) calibration and distributional validation of the misalignment generative model. (a) Expected-improvement (EI) search trajectory through the hyperparameter design space over 100 BO iterations. (b) Uniform-space diagnostics after marginal transformation, comparing the experimental and synthetic $U_x$ and $U_y$ distributions (25th percentile, median, and 75th percentile; black reference lines). (c) Latent $Z$-space (Gaussian) distributions for $Z_x$ and $Z_y$, showing close agreement in density levels between the experimental and synthetic data. (d--e) Hexbin density maps of the experimental and synthetic joint distributions of planar tilts $Z_x$ and $Z_y$. (f) Difference map (synthetic minus experimental), highlighting improved agreement in the tails due to joint-tail treatment and inclusion of high-misalignment motif fibers.}
    \label{fig8}
\end{figure}

The Bayesian optimization (BO) results after 100 iterations are summarized in Fig.~\ref{fig8}, which reports both the optimizer's search behavior and the resulting agreement between the experimental and synthetic datasets in the transformed uniform space and the corresponding latent Gaussian space. Figure~\ref{fig8}(a) shows the trajectory of the expected-improvement (EI) acquisition function through the design space; each iteration corresponds to a full synthetic reconstruction of the sub-volume, i.e., generating all fibers across the depth and evaluating the objective. After 100 iterations, the best solution achieved a total loss of approximately 12.6\%. The corresponding optimal hyperparameters and the individual loss components are reported in the Supplementary Material (Section 4).

To assess the quality of the marginal transformations, Fig.~\ref{fig8}(b) presents the uniform-space diagnostics. The synthetic and experimental transformed marginals are nearly identical. For $U_x$, the experimental percentiles are $P_{25}=0.24990$ and $P_{75}=0.75010$, while the synthetic counterpart yields $P_{25}=0.25031$ and $P_{75}=0.75010$. The same level of agreement is observed for $U_y$.

Figure~\ref{fig8}(c) compares the latent $Z$-space distributions and further confirms close agreement between the original and synthetic datasets. The maximum probability density is 0.408 for the experimental latent variables, $Z_X$ and $Z_Y$, and 0.409--0.407 for the synthetic counterpart, indicating that the calibrated model reproduces the dominant density level of the latent representation.

Finally, the dependence structure between the planar tilts is evaluated in Fig.~\ref{fig8}(d--f). ~\ref{fig8}(d) and \ref{fig8}(e) show the experimental and synthetic joint distributions of $Z_x$ and $Z_y$ using hexbin density maps, while Fig.~\ref{fig8}(f) plots the synthetic--experimental difference. The remaining discrepancy is minimal, on the order of $\sim 4\times10^{-4}$, and the improvement in the extremes highlights the role of explicitly treating joint-tail behavior beyond Gaussian tail independence and re-injecting high-misalignment motif fibers, thereby enabling agreement not only around the median but also in rare-event regions.

\begin{figure}[H]
    \centering
    \includegraphics[width=\textwidth]{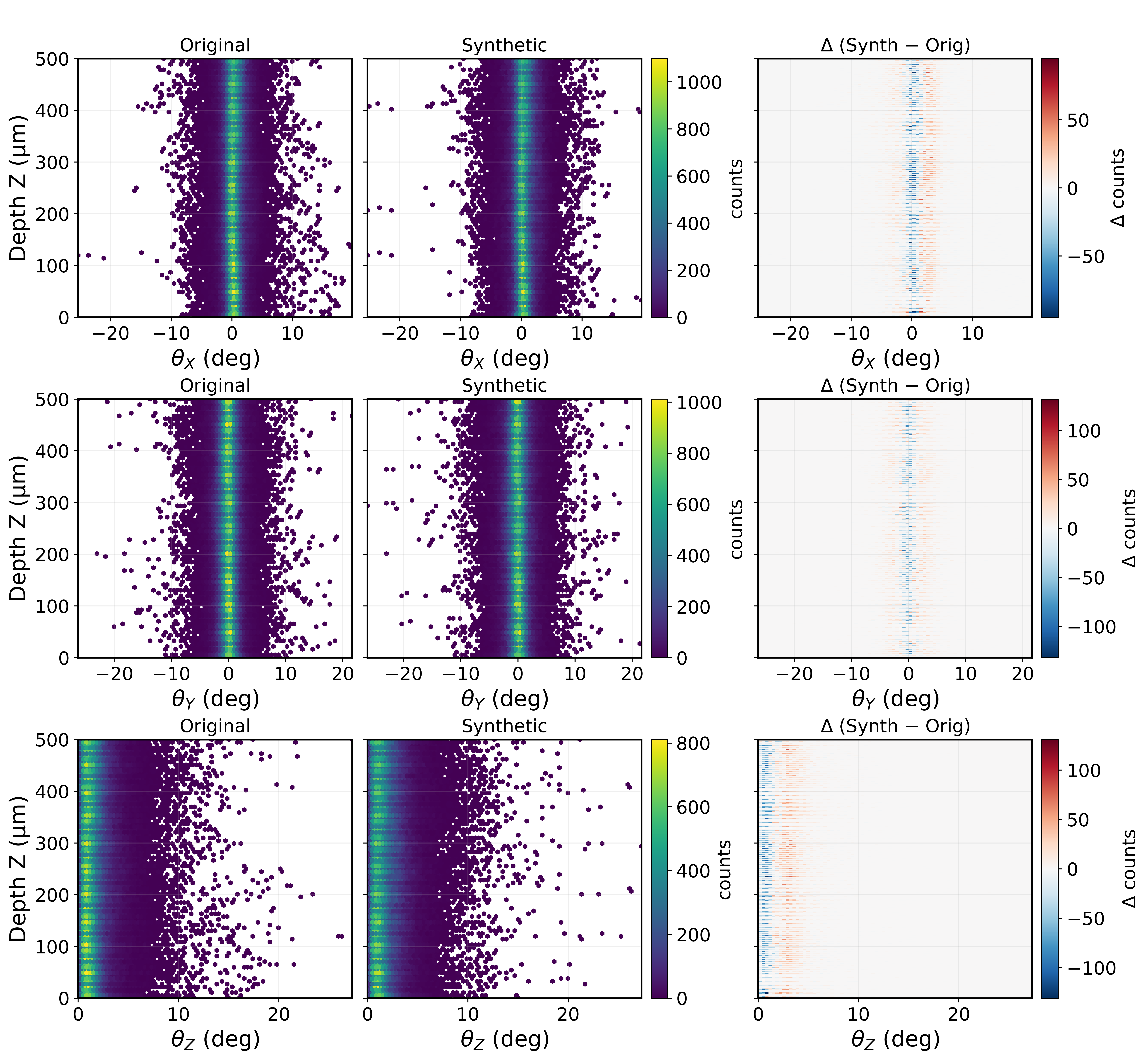}
    \caption{Depth-resolved hexbin density maps of the per-slice misalignment angles, highlighting agreement in the most probable depth-dependent trajectory and the spread of extreme events: original microstructure (left), synthetic microstructure (center), and synthetic--original density difference (right) for (a) $\theta_X$, (b) $\theta_Y$, and (c) $\theta_Z$.}
    \label{fig9}
\end{figure}

Furthermore, by concatenating the per-slice misalignment angles $\theta_X$, $\theta_Y$, and $\theta_Z$ across the full depth, the slice-wise information can be visualized as a two-dimensional density panel (angle versus depth) using hexbin maps. This representation captures both the most probable misalignment path, i.e., the mode of the distribution, and the statistical scatter of physically occurring angles at each depth. Figure~\ref{fig9} presents these results: Fig.~\ref{fig9}(a--c) show the experimental distributions, the synthetic statistically equivalent distributions, and the synthetic--experimental difference for $\theta_X$, $\theta_Y$, and $\theta_Z$, respectively.

The results demonstrate that the proposed model reproduces the depth-wise evolution of the dominant misalignment behavior, accurately tracking the path of the highest-density occurrences for all three angles while also retaining sufficient stochastic variability. Rare high-misalignment events are also represented. The residual discrepancy between the synthetic and experimental density is small, with a maximum per-slice absolute count differences of less than $4.17\%$ of the accumulated angles.

This figure is particularly important because it evaluates more than marginal agreement at isolated slices. It directly tests whether the model reproduces: (i) the most frequent depth-dependent trajectory of misalignment, (ii) distributional shape effects such as skewness and spread, and (iii) the presence and magnitude of extremes as a function of depth. It therefore provides a stringent and physically interpretable validation of statistical equivalence between the synthetic and original microstructures.

\begin{figure}[H]
    \centering
    \includegraphics[width=\textwidth]{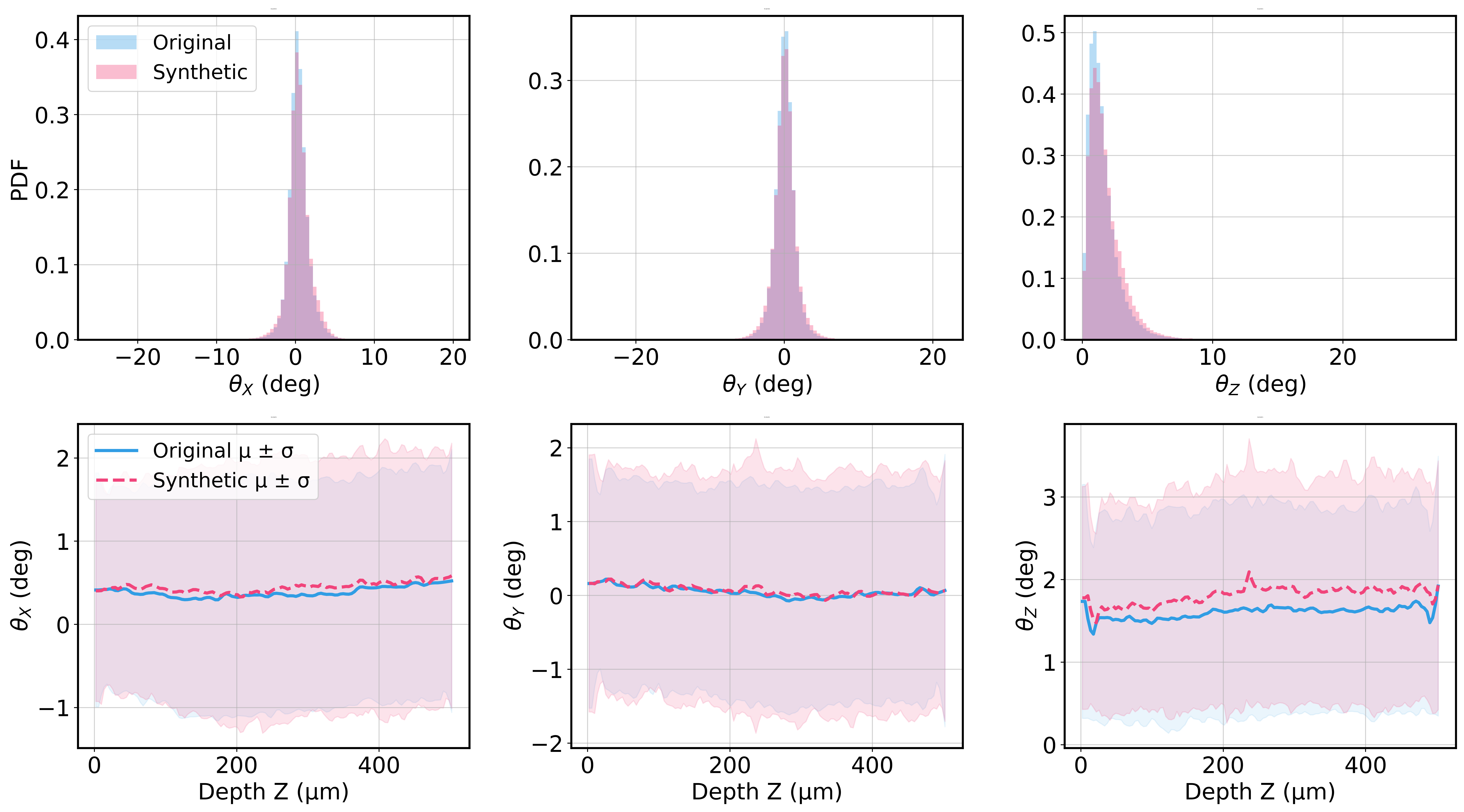}
    \caption{Global and depth-resolved statistical comparison of the original and synthetic misalignment distributions: (a) global PDF histograms for $\theta_X$, $\theta_Y$, and $\theta_Z$; (b) depth-dependent $\mu(z)\pm\sigma(z)$ envelopes.}
    \label{fig10}
\end{figure}

Figure~\ref{fig10} compares the global and depth-resolved misalignment statistics between the experimental (original) and generated (synthetic) datasets for $\theta_X$, $\theta_Y$, and $\theta_Z$. The global probability density functions shown in Fig.~\ref{fig10}(a) indicate close distributional agreement, as quantified by the two-sample KS statistic. The smallest discrepancy is observed for $\theta_Y$ ($D = 0.0229$), followed by $\theta_X$ ($D = 0.0328$), while $\theta_Z$ shows the largest, but still moderate, difference ($D = 0.0747$).

In addition to distributional similarity, the full support of the angles is largely preserved, and the peak density levels remain essentially unchanged. For example, the maximum PDF for $\theta_X$ differs only marginally between the experimental and synthetic datasets (0.411 versus 0.383 over their respective ranges). Similar behavior is found for $\theta_Y$, where both the minimum nonzero densities and peak densities are closely matched (maximum PDF: 0.357 for the original versus 0.336 for the synthetic dataset). For $\theta_Z$, the synthetic distribution exhibits a lower maximum PDF (0.443 versus 0.502), consistent with a reduced concentration near the dominant inclination region.

Figure~\ref{fig10}(b) evaluates whether the synthetic model reproduces the depth-wise evolution of misalignment. The maximum absolute difference between the $\mu(z)$ curves is small for $\theta_X$ and $\theta_Y$, reaching $0.133^\circ$ at $z = 281.250~\mu\text{m}$ and $0.107^\circ$ at $z = 238.281~\mu\text{m}$, respectively, confirming that the synthetic generator accurately tracks the depth-dependent trend of the planar tilts. For $\theta_Z$, the maximum mean deviation is larger, at $0.460^\circ$ at $z = 234.375~\mu\text{m}$, although it remains limited in magnitude relative to the overall depth profile. Importantly, the variability band is also well preserved: the $\mu \pm \sigma$ envelope span differs by only $8.406\%$ for $\theta_X$ and $8.118\%$ for $\theta_Y$, and remains within $2.675\%$ for $\theta_Z$. Collectively, these results demonstrate that the proposed model not only closely matches the global marginals and dependence structure, but also captures the depth-resolved evolution of both the central tendency and the dispersion.

\subsection{Synthetic microstructure generation}

\begin{figure}[H]
    \centering
    \includegraphics[width=\textwidth]{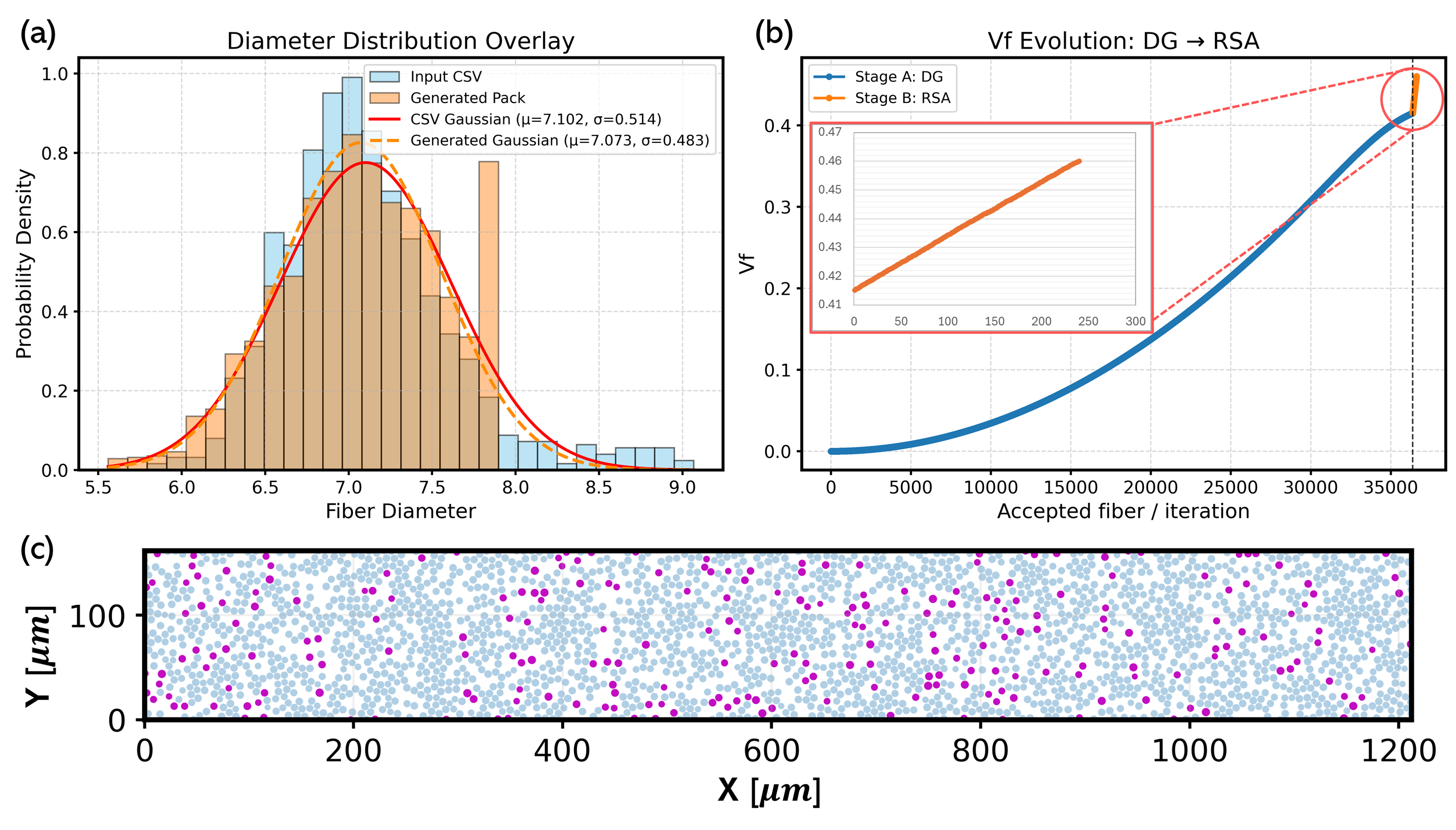}
    \caption{(a) Input and generated fiber-diameter distributions for the seed layer. (b) Fiber volume fraction evolution during the DG and RSA stages. (c) Final generated seed-layer microstructure (blue: DG; pink: RSA) with 2395 fibers and 46\% volume fraction.}
    \label{fig11}
\end{figure}

The initial seed layer and the fiber microstructure generation are shown in Fig.~\ref{fig11}, following the procedure described in Section~2.4.1. Figure~\ref{fig11}(a) compares the diameter distribution of the generated fibers with the input fiber-diameter PDF. The agreement is close, with the input dataset giving a fitted Gaussian mean diameter of $\mu = 7.1016$ and standard deviation $\sigma = 0.5144$, while the generated packing gives $\mu = 7.0734$ and $\sigma = 0.4830$. The generated synthetic seed microstructure was capped at $\mu - 3\sigma$ and $\mu + 1.5\sigma$ to limit excessively large fiber diameters, which are uncommon in aerospace-grade CFRPs. This explains why the generated packing does not contain very large fiber diameters. Figure~\ref{fig11}(b) shows the fiber volume fraction (VF) evolution during the dynamic growth (DG) and random sequential adsorption (RSA) stages. The DG stage exhibits a gradual exponential increase, reflecting the incremental growth of the fibers toward their target diameters, whereas the RSA stage fills the remaining gaps through random insertion at a faster rate. The final generated microstructure is presented in Fig.~\ref{fig11}(c), with a fiber count comparable to the experimental dataset (2395 fibers) and a final VF of 46\%. The seed layer microstructure packing required around 23~min of HPC run time with the same configuration described earlier in Section 2. 

\begin{figure}[H]
    \centering
    \includegraphics[width=\textwidth]{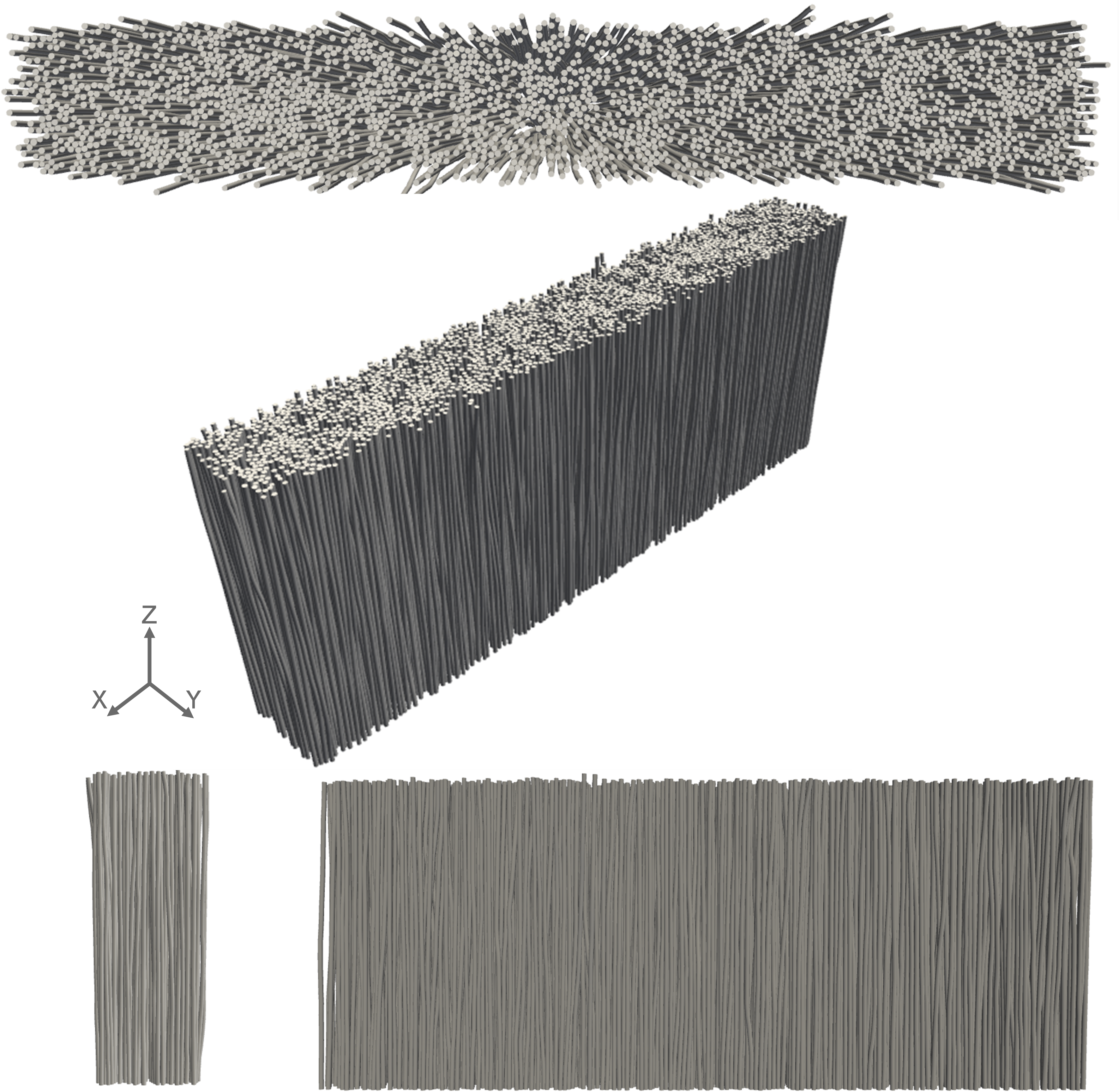}
    \caption{Final augmented synthetic microstructure generated by slice-by-slice 3D growth informed by the statistically equivalent modelling layer.}
    \label{fig12}
\end{figure}

Figure~\ref{fig12} shows the final augmented synthetic microstructure generated through slice-by-slice three-dimensional growth using the statistical modelling layer and the developed overlap-resolution solver described in Section~2.4.2. While the original X-ray-$\mu$CT data provide only fiber centerline information, the present augmentation enables reconstruction of a more realistic three-dimensional misaligned fiber microstructure with physically consistent radii, overlap-free geometry, and a physically admissible fiber network informed by a statistically equivalent modelling layer.

\section{Conclusion}

An integrated pipeline has been developed to transform X-ray-$\mu$CT fiber-path data into statistically equivalent and physically admissible synthetic three-dimensional microstructures. The framework combines slice-wise fiber misalignment quantification, stochastic statistical modelling, and overlap-free geometric generation within a unified workflow, thereby addressing a key limitation in current microstructure-informed modelling: the absence of a direct route from experimental imaging to realistic, simulation-ready synthetic media. The resulting microstructures retain the essential statistics of the original material while enhancing physical realism through consistent fiber radii and admissible fiber arrangements throughout the synthetic medium. This makes the approach valuable for virtual testing, micromechanics, and process-aware composite design, facilitating the synthetic generation of statistically representative and realistic three-dimensional fiber microstructures.

In future work, the framework will be extended to larger fiber domains, including large-scale 12K tow reconstruction, and the generated synthetic microstructures will be further validated against the original raw X-ray-$\mu$CT dataset. The framework will also be enriched with additional local descriptors, such as porosity, tortuosity, fiber conglomerates, and related morphological interactions, to enhance realism and predictive capability. Moreover, the stochastic consistency of the model will be evaluated through the reproducibility of multiple independently generated synthetic microstructures. More broadly, the framework provides the basis for an autonomous virtual laboratory of realistic fiber architectures and is readily extendable to different classes of fiber-reinforced polymer composites and other heterogeneous synthetic materials, including short-fiber systems.


\newpage

\section*{CRediT authorship contribution statement}

\textbf{Mohamad A. Raja}: Writing -- original draft, Writing -- review \& editing, Data curation, Software, Validation, Methodology, Investigation, Formal analysis, Conceptualization, Visualization. \textbf{Clemens Dransfeld}: Writing -- review \& editing, Supervision. \textbf{Boyang Chen}: Writing -- review \& editing, Supervision, Methodology, Conceptualization.

\section*{Declaration of Competing Interest}

The authors declare that they have no known competing financial interests or personal relationships that could have appeared to influence the work reported in this paper.

\section*{Acknowledgements}

The authors gratefully acknowledge the financial support provided by the Luchtvaart in Transitie (LiT) project. The authors also thank Dr.\ S.~Gomarasca for providing the raw X-ray-$\mu$CT datasets and for her support.

\section*{Appendix A. Supplementary data}

Supplementary material associated with this article is provided in a separate document, including additional details on the ellipse-intersection method, synthetic fiber modelling, microstructure generation, and supplementary results.

\section*{Data availability}

The data and associated codes form part of ongoing research and are therefore not currently publicly available. They will be made publicly available upon completion of the related studies through the author's website (https://www.mohamadraja.com/) and GitHub repository (https://github.com/MohamadA-RJ).

\newpage
 \bibliographystyle{elsarticle-num} 
 \bibliography{References}

\end{document}


\begin{frontmatter}

\title{\makebox[\textwidth][l]{Supplementary Information}\\[0.5em]
Autoregressive Modelling and Synthetic Generation of High-Fidelity, Statistically Equivalent 3D Microstructures for As-Manufactured Misalignments in Fiber-Reinforced Composites}

\author[tud]{Mohamad A. Raja\corref{cor1}}
\ead{M.A.Raja@tudelft.nl}

\author[tud]{Clemens Dransfeld}

\author[tud]{Boyang Chen\corref{cor2}}
\ead{b.chen-2@tudelft.nl}

\cortext[cor1]{Corresponding author. Address: Kluyverweg 1, 2629 HS Delft, The Netherlands. Office: Building 62, NB0.42.}
\cortext[cor2]{Corresponding author. Address: Kluyverweg 1, 2629 HS Delft, The Netherlands. Office: Building 62, NB0.37.}
\affiliation[tud]{organization={Delft University of Technology (TU Delft), Faculty of Aerospace Engineering, Department of Aerospace Structures and Materials},
    addressline={Kluyverweg 1},
    city={Delft},
    postcode={2629 HS},
    country={The Netherlands}}

\end{frontmatter}

\newpage

\begin{figure}[H]
    \centering
    \includegraphics[width=\textwidth]{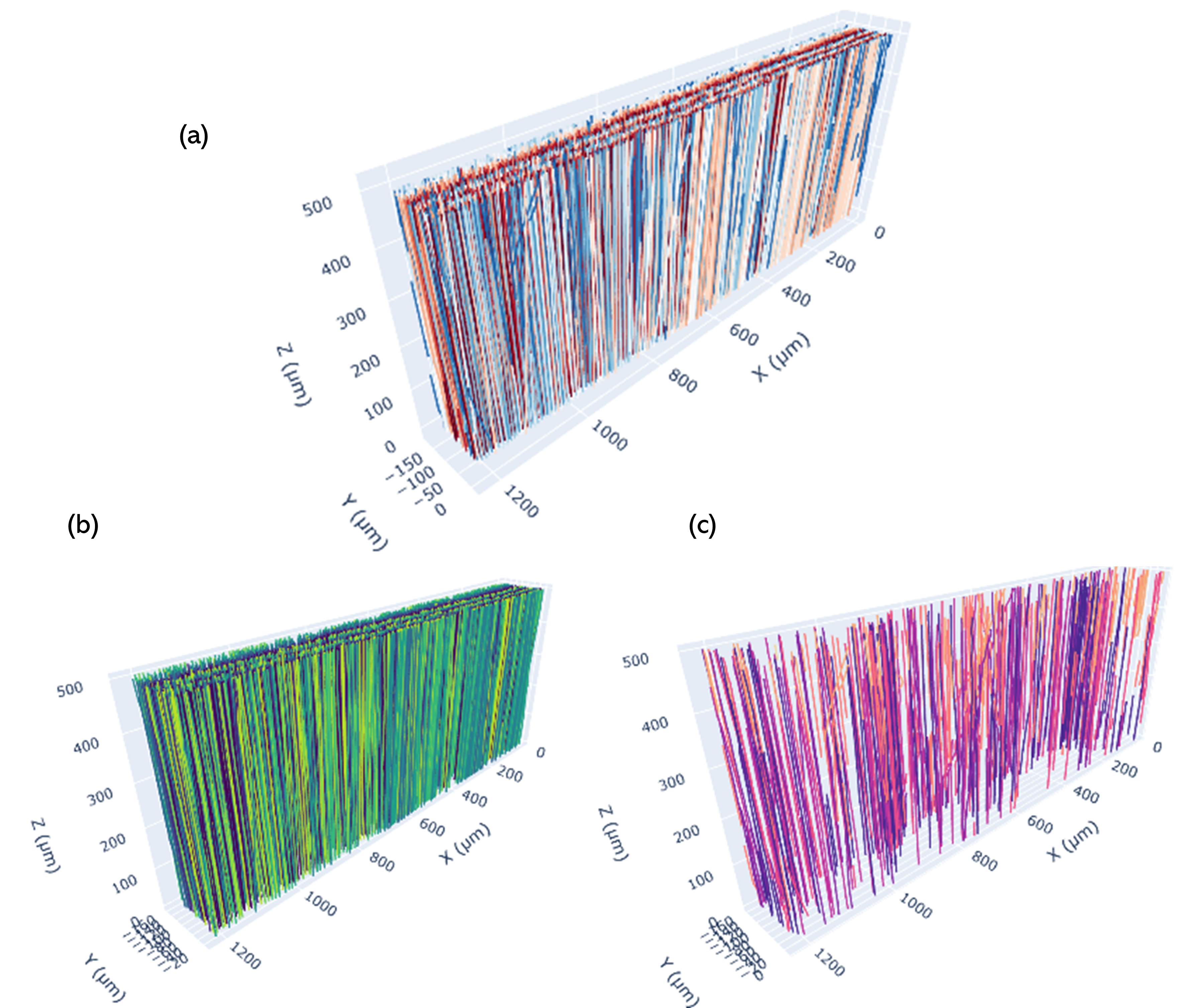}
    \caption{Xray-$\mu$CT experimental sub-volume showing the full segmented fiber population, including continuous and non-continuous fibers (2890 fibers). (b) Extracted continuous fibers (2395 fibers). (c) Extracted non-continuous fibers (495 fibers), illustrating their truncated trajectories due to limited reconstructed depth.}
    \label{figS1}
\end{figure}

\newpage

\section{Ellipse-intersection method: formulation and parametrization}

This supplementary section provides the full derivation and supporting details for the ellipse--intersection approach used to analyze fiber misalignment from three-dimensional X-ray-$\mu$CT centerlines. We include: (i) the local cylinder model and its surface parameterization, (ii) construction of a robust orthonormal cross-sectional frame, (iii) derivation of ellipse parameters from the cylinder--plane intersection, and (iv) the equivalence between the vector and rotation-matrix forms of the ellipse.

\subsection{Cylinder surface parameterization}

This formulation is used for rendering and visualization. A point on the cylinder axis is expressed as
\begin{equation}
\mathbf{c}(s)=\mathbf{p}_0+s\,\mathbf{u},
\qquad
s\in[0,L]
\label{eq:S1}
\end{equation}

At each $s$, the cross-section lies in the plane orthogonal to $\mathbf{u}$. Let $\{\mathbf{e}_x,\mathbf{e}_y\}$ be an orthonormal basis spanning that plane. A circle of radius $R$ in that plane is
\begin{equation}
\mathbf{r}(\theta)=R\left(\cos\theta\,\mathbf{e}_x+\sin\theta\,\mathbf{e}_y\right),
\qquad
\theta\in[0,2\pi)
\label{eq:S2}
\end{equation}

Sweeping the circle along the axis yields the cylinder surface:
\begin{equation}
\mathbf{S}(s,\theta)
=
\mathbf{p}_0+s\,\mathbf{u}
+
R\cos\theta\,\mathbf{e}_x
+
R\sin\theta\,\mathbf{e}_y
\label{eq:S3}
\end{equation}
where $s\in[0,L]$ and $\theta\in[0,2\pi)$. For a fixed $s$, $\mathbf{S}(s,\theta)-\mathbf{c}(s)=\mathbf{r}(\theta)$ lies in the plane normal to $\mathbf{u}$, and its norm is $R$ because $\mathbf{e}_x$ and $\mathbf{e}_y$ are orthonormal. Varying $s$ traces the straight axis $\mathbf{c}(s)$; hence, $\mathbf{S}(s,\theta)$ describes a right circular cylinder.

\subsection{Constructing a robust local cross-sectional frame}

To define $\mathbf{e}_x$ and $\mathbf{e}_y$ robustly, a seed vector $\mathbf{t}$ must not be nearly colinear with $\mathbf{u}$. The canonical axis least aligned with $\mathbf{u}$ is selected as
\begin{equation}
\mathbf{t}=
\begin{cases}
(1,0,0), & \text{if } |u_x|\le |u_y| \text{ and } |u_x|\le |u_z|,\\
(0,1,0), & \text{if } |u_y|\le |u_x| \text{ and } |u_y|\le |u_z|,\\
(0,0,1), & \text{otherwise.}
\end{cases}
\label{eq:S4}
\end{equation}

Then define
\[
\mathbf{e}_x=\frac{\mathbf{u}\times\mathbf{t}}{\lVert\mathbf{u}\times\mathbf{t}\rVert},
\qquad
\mathbf{e}_y=\mathbf{u}\times\mathbf{e}_x .
\]
These satisfy
\[
\mathbf{e}_x\cdot\mathbf{u}=0,
\qquad
\mathbf{e}_y\cdot\mathbf{u}=0,
\qquad
\mathbf{e}_x\cdot\mathbf{e}_y=0,
\qquad
\lVert\mathbf{e}_x\rVert=\lVert\mathbf{e}_y\rVert=1.
\]
Hence, $\{\mathbf{e}_x,\mathbf{e}_y,\mathbf{u}\}$ forms a right-handed orthonormal frame.

If $\lVert\mathbf{u}\times\mathbf{t}\rVert$ falls below a small tolerance, indicating a near-parallel case, a different $\mathbf{t}$ is selected. The frame is built per segment, which may introduce a mild ring twist for visualization. Importantly, the ellipse parameters at a slice, $(a,b,\varphi)$, are invariant to this choice; only the visual seam orientation is affected.

\subsection{Ellipse orientation and semi-axes from the local segment direction}

Let the in-plane projection of the axis be $\mathbf{g}=(u_x,u_y)$. For $u_x^2+u_y^2>0$, define the unit in-plane direction and its perpendicular as
\begin{equation}
\mathbf{g}
=
\frac{(u_x,u_y)}{\sqrt{u_x^2+u_y^2}},
\qquad
\mathbf{g}_{\perp}=(-g_y,g_x)
\label{eq:S5}
\end{equation}

The in-plane azimuth of the major axis is
\begin{equation}
\varphi=\operatorname{atan2}(u_y,u_x)
\label{eq:S6}
\end{equation}

For a right circular cylinder of radius $R$ intersected by a horizontal plane, the intersection is an ellipse with semi-axes
\begin{equation}
a=\frac{R}{|u_z|},
\qquad
b=R
\label{eq:S7}
\end{equation}
where $a$ is the major semi-axis aligned with $\mathbf{g}$, and $b$ is the minor semi-axis aligned with $\mathbf{g}_{\perp}$. The major axis length increases as $|u_z|$ decreases, i.e., as the segment becomes more tilted relative to the slice normal.

In the limiting case of a vertical segment, if $u_x\approx u_y\approx 0$, then $u_z\approx \pm 1$ and $a\approx b\approx R$, so the ellipse degenerates to a circle; in that case, $\varphi$ is undefined but unnecessary.

\subsection{Parametric ellipse in global slice coordinates}

In vector form, using $\mathbf{c}_{xy}$, $\mathbf{g}$, and $\mathbf{g}_{\perp}$, the ellipse in the global $xy$-plane is
\begin{equation}
\mathbf{x}(\theta)
=
\mathbf{c}_{xy}
+
a\cos\theta\,\mathbf{g}
+
b\sin\theta\,\mathbf{g}_{\perp},
\qquad
\theta\in[0,2\pi)
\label{eq:S8}
\end{equation}
This form is convenient for generating discrete ellipse points for plotting and for subsequent packing and overlap checks.

Equivalently, the same result can be derived using a rotation matrix applied to an ellipse centered at the origin. Start with an axis-aligned ellipse:
\begin{equation}
x'(\theta)=a\cos\theta,
\qquad
y'(\theta)=b\sin\theta
\label{eq:S9}
\end{equation}

Rotate by the in-plane angle $\varphi$:
\begin{equation}
R_{\varphi}
=
\begin{bmatrix}
\cos\varphi & -\sin\varphi\\
\sin\varphi & \cos\varphi
\end{bmatrix}
\label{eq:S10}
\end{equation}

Then
\begin{equation}
\begin{bmatrix}
x_r(\theta)\\
y_r(\theta)
\end{bmatrix}
=
R_{\varphi}
\begin{bmatrix}
a\cos\theta\\
b\sin\theta
\end{bmatrix}
=
\begin{bmatrix}
a\cos\theta\cos\varphi-b\sin\theta\sin\varphi\\
a\cos\theta\sin\varphi+b\sin\theta\cos\varphi
\end{bmatrix}
\label{eq:S11}
\end{equation}

Finally, translate to the ellipse center $\mathbf{c}=(c_x,c_y)$:
\begin{equation}
x(\theta)=c_x+a\cos\theta\cos\varphi-b\sin\theta\sin\varphi
\label{eq:S12}
\end{equation}
\begin{equation}
y(\theta)=c_y+a\cos\theta\sin\varphi+b\sin\theta\cos\varphi
\label{eq:S13}
\end{equation}

For visualization, the cylinder surface can be discretized by sampling $s$ into $N_c$ rings along $[0,L]$ and $\theta$ into $N_t$ points per ring over $[0,2\pi)$. Evaluating $\mathbf{S}(s,\theta)$ on this grid, the choice of $\mathbf{e}_x$ and $\mathbf{e}_y$ rotates the rings about the axis but does not affect the geometric cylinder nor the ellipse parameters at slice intersections.

\subsection*{Proof 1. Relation of depth misalignment between the ellipse method and the CDM}

Let
\[
p=\frac{dX}{dZ},
\qquad
q=\frac{dY}{dZ}.
\]
A non-unit tangent is proportional to $(p,q,1)$, so after normalization:
\begin{equation}
\mathbf{u}
=
\frac{(p,q,1)}{\sqrt{1+p^2+q^2}}
\label{eq:S14}
\end{equation}
and therefore
\begin{equation}
u_z=\frac{1}{\sqrt{1+p^2+q^2}}
\label{eq:S15}
\end{equation}

From the dot product with $\mathbf{e}_z$, and substituting Eq.~\eqref{eq:S15},
\begin{equation}
\theta_Z=\arccos\left(|u_z|\right)
\label{eq:S16}
\end{equation}
which becomes
\begin{equation}
\theta_Z
=
\arccos\left(
\frac{1}{\sqrt{1+p^2+q^2}}
\right)
\label{eq:S17}
\end{equation}

Now let
\[
r=\sqrt{p^2+q^2}.
\]
Using the identity
\[
\arccos\left(\frac{1}{\sqrt{1+r^2}}\right)=\arctan(r),
\]
we obtain
\begin{equation}
\theta_Z
=
\arctan\left(\sqrt{p^2+q^2}\right)
=
\arctan\left(
\sqrt{
\left(\frac{dX}{dZ}\right)^2
+
\left(\frac{dY}{dZ}\right)^2
}
\right)
\label{eq:S18}
\end{equation}

Equation~\eqref{eq:S16} is the form obtained from the ellipse-intersection method, whereas Eq.~\eqref{eq:S18} is the corresponding form obtained from the central difference method.

\newpage

\section{Synthetic fiber modelling: Gaussian copula, autoregressive model, tail-independence boosting, high-misalignment motifs, and Bayesian optimization}

\subsection*{Proof 2. Using Cholesky decomposition for the covariance matrix}

We aim to generate two correlated Gaussian latent variables with zero mean such that
\[
(Z_x,Z_y)\sim\mathcal{N}(0,\Sigma_{s}),
\]
with covariance matrix
\begin{equation}
\Sigma_{s}=
\begin{bmatrix}
1 & \rho_G\\
\rho_G & 1
\end{bmatrix}
\label{eq:S19}
\end{equation}

If one were to multiply the white-noise vector directly by $\Sigma_{s}$, i.e.,
\[
\mathrm{Z}_{s}=\Sigma_{s}\,\varepsilon_{s},
\]
where $\varepsilon_{s}$ is standard Gaussian white noise, the resulting covariance would not equal $\Sigma_{s}$. To show this, start from the general definition. For a random column vector $\mathbf{X}=[X_1,X_2,\ldots,X_n]^T$ with mean $\boldsymbol{\mu}=\mathop{\mathbb{E}}[\mathbf{X}]$, the covariance matrix is
\begin{equation}
\Sigma=\operatorname{Cov}(\mathbf{X})
=
\mathop{\mathbb{E}}\left[
(\mathbf{X}-\boldsymbol{\mu})(\mathbf{X}-\boldsymbol{\mu})^T
\right]
\label{eq:S20}
\end{equation}

The resulting $n\times n$ matrix has the form
\begin{equation}
\Sigma=
\begin{bmatrix}
\operatorname{Var}(X_1) & \operatorname{Cov}(X_1,X_2) & \cdots & \operatorname{Cov}(X_1,X_n)\\
\operatorname{Cov}(X_2,X_1) & \operatorname{Var}(X_2) & \cdots & \operatorname{Cov}(X_2,X_n)\\
\vdots & \vdots & \ddots & \vdots\\
\operatorname{Cov}(X_n,X_1) & \operatorname{Cov}(X_n,X_2) & \cdots & \operatorname{Var}(X_n)
\end{bmatrix}
\label{eq:S21}
\end{equation}

It should be noted that the covariance matrix is symmetric, i.e.,
\[
\operatorname{Cov}(X_i,X_j)=\operatorname{Cov}(X_j,X_i),
\]
hence $\Sigma=\Sigma^T$. The diagonal entries are the variances of each variable, $\Sigma_{ii}=\operatorname{Var}(X_i)$, while the off-diagonal entries measure the linear dependence between components. The matrix is always positive semi-definite.

Using Eq.~\eqref{eq:S20} for the generated latent vector $\mathrm{Z}_{s}=\Sigma_{s}\,\varepsilon_{s},$, one obtains
\begin{equation}
\operatorname{Cov}(\Sigma_{s}\varepsilon_{s})
=
\mathbb{E}\left[
(\Sigma_{s}\varepsilon_{s})(\Sigma_{s}\varepsilon_{s})^T
\right]
\label{eq:S22}
\end{equation}

Expanding the transpose and moving the constant matrices outside the expectation gives
\begin{equation}
\mathop{\mathbb{E}}\left[\Sigma_{s}\varepsilon_{s}(\varepsilon_{s})^T(\Sigma_{s})^T\right]
=
\Sigma_s\,\mathop{\mathbb{E}}\left[\varepsilon_s\varepsilon_s^T\right](\Sigma_{s})^T
\label{eq:S23}
\end{equation}

Since
\[
\mathop{\mathbb{E}}\left[\varepsilon_{s}(\varepsilon_{s})^T\right]=\operatorname{Cov}(\varepsilon_{s})=I_2,
\]
it follows that
\begin{equation}
\operatorname{Cov}({Z}_{s})
=
\operatorname{Cov}(\Sigma_{s}\,\varepsilon_{s})
=
\Sigma_{s} I_2(\Sigma_{s})^T
=
(\Sigma_{s})^2
\neq
\Sigma_{s}
\label{eq:S24}
\end{equation}

Thus, direct multiplication by $\Sigma_{s}$ over-scales the covariance. In general, for any random vector $\mathbf{X}$ and constant matrix $A$,
\[
\operatorname{Cov}(A\mathbf{X}) = A\,\operatorname{Cov}(\mathbf{X})\,A^T.
\]

Therefore, we need a matrix $L$ such that
\begin{equation}
\Sigma = LL^T
\label{eq:S25}
\end{equation}

Then
\begin{equation}
\operatorname{Cov}(\mathrm{Z_{s}})
=
\operatorname{Cov}(L_{s}\varepsilon_{s})
=
L_{s}\,\operatorname{Cov}(\varepsilon_{s})\,(L_{s})^T
=
L_{s} I_2 (L_{s})^T
=
L_{s}(L_{s})^T
=
\Sigma_{s}
\label{eq:S26}
\end{equation}

The matrix $L$ is obtained from the Cholesky decomposition of $\Sigma$. For a general $2\times 2$ symmetric matrix
\[
A=
\begin{bmatrix}
a_{11} & a_{12}\\
a_{12} & a_{22}
\end{bmatrix},
\]
the Cholesky factorization is $A=LL^T$, where
\begin{equation}
L=
\begin{bmatrix}
l_{11} & 0\\
l_{21} & l_{22}
\end{bmatrix}
\label{eq:S27}
\end{equation}
with
\[
l_{11}=\sqrt{a_{11}},
\qquad
l_{21}=\frac{a_{12}}{l_{11}},
\qquad
l_{22}=\sqrt{a_{22}-l_{21}^2}.
\]

Thus, for the per-slice covariance matrix, the Cholesky decomposition is
\begin{equation}
L=
\begin{bmatrix}
1 & 0\\
\rho_G & \sqrt{1-\rho_G^2}
\end{bmatrix}
\label{eq:S28}
\end{equation}
which guarantees
\begin{equation}
L_{s}(L_{s})^T=
\begin{bmatrix}
1 & \rho_G\\
\rho_G & 1
\end{bmatrix}
=
\Sigma_{s}
\label{eq:S29}
\end{equation}

\subsection{Implementation of the stationary autoregressive process}

To model fiber persistence across the depth, a stationary autoregressive (AR) model is implemented in the latent Gaussian space after sampling a latent vector from the bivariate distribution and applying the Cholesky decomposition described above. This concept is widely adopted in time-series forecasting.

A stationary time series is a stochastic process whose statistical properties do not depend on the time at which it is observed. In particular, the mean and covariance remain constant over time. For a stationary process $S(t)$, the mean is time-independent:
\begin{equation}
\mathbb{E}[S(t)] = \mathbb{E}[S_t] = \mu
\label{eq:S30}
\end{equation}

The covariance depends only on the lag $\tau$ and not on the absolute time $t$:
\begin{equation}
\operatorname{Cov}(S_t,S_{t-\tau})
=
\mathbb{E}\left[(S_t-\mu)(S_{t-\tau}-\mu)\right]
\label{eq:S31}
\end{equation}

Expanding gives
\begin{align}
\operatorname{Cov}(S_t,S_{t-\tau})
&=
\mathbb{E}\left[S_tS_{t-\tau}-\mu S_{t-\tau}-\mu S_t+\mu^2\right]
\label{eq:S32}\\
&=
\mathbb{E}[S_tS_{t-\tau}] - \mu \mathbb{E}[S_{t-\tau}] - \mu \mathbb{E}[S_t] + \mu^2
\label{eq:S33}\\
&=
\mathbb{E}[S_tS_{t-\tau}] - 2\mu^2 + \mu^2
\label{eq:S34}\\
&=
\mathbb{E}[S_tS_{t-\tau}] - \mu^2
\label{eq:S35}
\end{align}

The variance, corresponding to lag $\tau=0$, is therefore also constant:
\begin{equation}
\operatorname{Var}(S_t)
=
E\left[(S_t-E[S_t])^2\right]
=
E\left[(S_t-\mu)^2\right]
=
\sigma^2
\label{eq:S36}
\end{equation}

These conditions motivate the preprocessing step in which the misalignment marginals are transformed into standard Gaussian variables.

With this in mind, an AR model for a stationary stochastic process can be introduced. In a general AR process of order $p$, the state at time $t$ is written as
\begin{equation}
S_t = \sum_{i=1}^{p}\phi_i S_{t-i} + e_t
\label{eq:S37}
\end{equation}
or equivalently
\begin{equation}
S_t = \phi_1 S_{t-1} + \cdots + \phi_p S_{t-p} + e_t
\label{eq:S38}
\end{equation}
where $e_t$ is an uncorrelated white-noise term.

For a stationary zero-mean AR process,
\begin{equation}
\mathbb{E}[S_t]=0,
\qquad
\operatorname{Var}(S_t)=\sigma^2,
\qquad
\forall t
\label{eq:S39}
\end{equation}

In this work, a zero-mean first-order (i.e., lag-1) autoregressive process is used:
\begin{equation}
S_t = \phi_1 S_{t-1} + e_t
\label{eq:S40}
\end{equation}
where $e_t$ is i.i.d.\ noise distributed as $e_t\sim\mathcal{N}(0,\sigma^2)$.

This corresponds to the fiber case as
\begin{equation}
Z_s
=
\phi\,Z_{s-1}
+
\sqrt{1-\phi^2}\,Z_{\mathrm{noise}}^{(s)},
\qquad
Z_{\mathrm{noise}}^{(s)}=L^{(s)}\varepsilon_s,
\qquad
\varepsilon_s\sim\mathcal{N}(0,I_2)
\label{eq:S41}
\end{equation}
where $\phi$ is the AR coefficient that describes the persistence of fibers from one slice to the next. For AR(1) stationarity, $|\phi|<1$ is required. The factor $\sqrt{1-\phi^2}$ is the variance-correction term, as shown next for the univariate case.

\subsection*{Proof 3. Variance-correction term for the autoregressive model}

Let
\begin{equation}
Z_t = a Z_{t-1} + \eta_t,
\qquad
\eta_t\sim\mathcal{N}(0,\sigma^2)
\label{eq:S42}
\end{equation}
where $a$ is the autoregressive coefficient and $\eta_t$ is white noise. This process is required to satisfy
\begin{equation}
\mathbb{E}[Z_t]=0,
\qquad
\operatorname{Var}(Z_t)=1
\label{eq:S43}
\end{equation}

Taking expectations on both sides of Eq.~\eqref{eq:S42} gives
\begin{equation}
\mathbb{E}[Z_t]=a\mathbb{E}[Z_{t-1}]+\mathbb{E}[\eta_t]
\label{eq:S44}
\end{equation}
which is automatically satisfied since $\mathbb{E}[\eta_t]=0$ and $\mathbb{E}[Z_{t-1}]=0$.

Now consider the variance:
\begin{equation}
\operatorname{Var}(Z_t)
=
\mathbb{E}\left[(Z_t-E[Z_t])^2\right]
=
\mathbb{E}[Z_t^2]
\label{eq:S45}
\end{equation}

Expanding $Z_t^2$:
\begin{equation}
Z_t^2 = (aZ_{t-1}+\eta_t)^2 = a^2 Z_{t-1}^2 + 2aZ_{t-1}\eta_t + \eta_t^2
\label{eq:S46}
\end{equation}

Taking expectations:
\begin{equation}
\mathbb{E}[Z_t^2]
=
a^2\mathbb{E}[Z_{t-1}^2]
+
2a\mathbb{E}[Z_{t-1}\eta_t]
+
\mathbb{E}[\eta_t^2]
\label{eq:S47}
\end{equation}

Since $Z_{t-1}$ and $\eta_t$ are independent and $E[\eta_t]=0$,
\[
E[Z_{t-1}\eta_t]=E[Z_{t-1}]E[\eta_t]=0.
\]
Therefore,
\begin{equation}
\operatorname{Var}(Z_t)
=
a^2\operatorname{Var}(Z_{t-1})
+
\operatorname{Var}(\eta_t)
\label{eq:S48}
\end{equation}

Imposing $\operatorname{Var}(Z_t)=\operatorname{Var}(Z_{t-1})=1$ gives
\begin{equation}
1 = a^2\cdot 1 + \operatorname{Var}(\eta_t)
\quad\Rightarrow\quad
\operatorname{Var}(\eta_t)=\sigma^2=1-a^2
\label{eq:S49}
\end{equation}

Since scaling a random variable by a constant scales its variance by the square of that constant,
\begin{equation}
\operatorname{Var}(a\eta_t)=a^2\operatorname{Var}(\eta_t)
\label{eq:S50}
\end{equation}
the variance-correction term must therefore be
\[
\sigma=\sqrt{1-a^2}.
\]

Hence the corrected recursion is
\begin{equation}
Z_t = a Z_{t-1} + \sigma\eta_t
\label{eq:S51}
\end{equation}

If $\eta_t$ is instead taken as standard white noise with $\operatorname{Var}(\eta_t)=1$, then the variance-preserving form is
\begin{equation}
Z_t = aZ_{t-1} + \sigma\eta_t
\label{eq:S52}
\end{equation}
with $\sigma=\sqrt{1-a^2}$.

Applying the variance operator:
\begin{equation}
\operatorname{Var}(Z_t)
=
\operatorname{Var}(aZ_{t-1}+\sigma\eta_t)
\label{eq:S53}
\end{equation}

Using independence between $Z_{t-1}$ and $\eta_t$,
\begin{equation}
\operatorname{Var}(Z_t)
=
\operatorname{Var}(aZ_{t-1})
+
\operatorname{Var}(\sigma\eta_t)
\label{eq:S54}
\end{equation}

Pulling out constants,
\begin{equation}
\operatorname{Var}(Z_t)
=
a^2\operatorname{Var}(Z_{t-1})
+
\sigma^2\operatorname{Var}(\eta_t)
\label{eq:S55}
\end{equation}

Since $\operatorname{Var}(Z_t)=\operatorname{Var}(Z_{t-1})=\operatorname{Var}(\eta_t)=1$, it follows that
\begin{equation}
\sigma=\sqrt{1-a^2}
\label{eq:S56}
\end{equation}

This proves that the scalar multiplied with the noise in Eq.~\eqref{eq:S52} must be $\sqrt{1-a^2}$. This makes sure that the individual components of $Z_s$ stay as standard normal. 

So far, the derivation has been presented in the univariate setting for simplicity. The same reasoning applies to the multivariate case used here, where $Z_s\in\mathbb{R}^2$ and the target covariance is
\[
\Sigma_{s} = L_{s}L_{s}^T.
\]

The vector recursion is
\begin{equation}
Z_s
=
\phiZ_{s-1}
+
\sqrt{1-(\phi)^2}\,\left(L_{s}\varepsilon_s\right)
\label{eq:S57}
\end{equation}

Its covariance satisfies
\begin{equation}
\operatorname{Cov}(Z_s)
=
\operatorname{Cov}
\left(
\phi~Z_{s-1}
+
\sqrt{1-(\phi)^2}\,L_{s}\varepsilon_s
\right)
\label{eq:S58}
\end{equation}

Since $\varepsilon_s$ and $Z_{s-1}$ are independent,
\begin{equation}
\operatorname{Cov}(Z_s)
=
(\phi)^2\operatorname{Cov}(Z_{s-1})
+
(1-(\phi)^2)\operatorname{Cov}(L_{s}\varepsilon_s)
\label{eq:S59}
\end{equation}

Since,
\[
\operatorname{Cov}(L_{s}\varepsilon_s)
=
L_{s}\operatorname{Cov}(\varepsilon_s)L_{s}^T
=
L_{s}IL_{s}^T
=
\Sigma_{s}.
\]
Therefore,
\begin{equation}
\operatorname{Cov}(Z_s)
=
\phi^2~\Sigma_{s-1}
+
(1-\phi^2)~\Sigma_{s}
\label{eq:S60}
\end{equation}

\noindent The deviation from exact marginal preservation is
\begin{equation}
\operatorname{Cov}(Z_s) - \Sigma_s
=
\phi^2\!~\left(\Sigma_{s-1} - \Sigma_s\right),
\label{eq:S61}
\end{equation}
which is proportional to both the inter-slice covariance gradient and the square
of the persistence coefficient.  For the fibrous microstructures studied here,
adjacent-slice covariances vary smoothly and the differences
$\|\Sigma_{s-1} - \Sigma_s\|$ are small relative to $\|\Sigma_s\|$,
so the deviation is correspondingly small.  Furthermore,
since $0 \leq \phi_{\mathrm{global}} < 1$, the prefactor $(\phi)^2$ is strictly less than $\phi_{\mathrm{global}}$ itself, providing an inherent additional attenuation of the error term beyond what the raw persistence coefficient would suggest. The AR(1) update therefore produces a natural, physically meaningful smoothing of the copula structure across depth, rather than an abrupt slice-to-slice jump, and this behaviour mirrors the physical continuity of fibres across the microstructure volume with satisfactory results, especially after Bayesian optimization process. The exact enforcement of the covariance for fiber persistence modelling is left for future work.

\subsection{Regarding the lag-1 autocorrelation coefficient}
In a continuous Ornstein--Uhlenbeck (OU) process, the correlation decays exponentially with distance. Upon discretization, the smoothness should remain consistent regardless of the step size $\Delta z$.

The autocorrelation of an OU process at distance $\Delta z$ is
\begin{equation}
\rho(\Delta z)=e^{-\theta\Delta z}
\label{eq:S62}
\end{equation}
where $\theta$ is the 'forgetting' rate.

A reference correlation $\phi_{\mathrm{global}}$ is defined at a reference spacing $\Delta z_{\mathrm{ref}}$:
\begin{equation}
\phi_{\mathrm{global}}=e^{-\theta \Delta z_{\mathrm{ref}}}
\label{eq:S63}
\end{equation}

Solving for $\theta$ gives
\begin{equation}
\ln(\phi_{\mathrm{global}})
=
-\theta \Delta z_{\mathrm{ref}}
\quad\Rightarrow\quad
\theta
=
-\frac{\ln\phi}{\Delta z_{\mathrm{ref}}}
\label{eq:S64}
\end{equation}

The step-wise AR coefficient for a simulation spacing $\Delta z$ is then
\begin{equation}
a
=
e^{-\theta\Delta z}
=
e^{\left(\frac{\ln\phi}{\Delta z_{\mathrm{ref}}}\right)\Delta z}
\label{eq:S65}
\end{equation}

Using the identity $e^{\ln(x)y}=x^y$, this simplifies to
\begin{equation}
a=\phi_{\mathrm{global}}^{\Delta z/\Delta z_{\mathrm{ref}}}
\label{eq:S66}
\end{equation}

In this work, $\Delta z_{\mathrm{ref}}=\Delta z_{\mathrm{slice}}=1$, since both the
statistical modelling and the microstructure generation operate at the same
discretisation: one cross-sectional slice per step.  The ratio therefore reduces to
unity and Eq.~\eqref{eq:S66} simplifies to
\[
a=\phi.
\]
In the model, $a$ is the lag-1 autocorrelation coefficient, i.e., the correlation between $Z_s$ and $Z_{s-1}$.

While $\phi$ characterises the persistence of the microstructure at the sub-volume level, encoding the average depth-wise straightness of fibres across the entire volume, individual fibres in a real composite do not all share identical persistence as manufacturing variability and other factors introduce fibre-to-fibre differences in curvature continuity. To capture this physical variability, each fibre $i$ is assigned its own persistence coefficient drawn independently from a Gaussian distribution centred on the global value:
\begin{equation}
\phi
=
\phi_{\mathrm{global}}
+
\sigma_{\phi}\,\eta_i,
\qquad
\eta_i \sim \mathcal{N}(0,1),
\label{eq:S67}
\end{equation}

where $\sigma_{\phi}$ is the \emph{jitter} standard deviation. This means that $\phi = \phi_{\mathrm{local},i}$, as the lag-1 coefficient used in the AR(1) recursion, is therefore $a = \phi_{\mathrm{local},i}$ for each fibre rather than the fixed global value. This two-parameter representation $(\phi_{\mathrm{global}},\,\sigma_{\phi})$ is more expressive than a single shared coefficient: $\phi_{\mathrm{global}}$ controls the ensemble-mean smoothness of all fibres, while $\sigma_{\phi}$ controls the spread of individual fibre behaviors around that mean, producing a realistic population of fibres with varying degrees of curvature.

Both $\phi_{\mathrm{global}}$ and $\sigma_{\phi}$ are treated as free parameters and jointly optimized by the Bayesian optimization procedure.  The optimizer identifies the combination that best reproduces the empirical model hyperparamters close to the reference microstructure, ensuring that the per-fibre persistence variability is calibrated to the observed data rather than prescribed by assumption.

\subsection{Joint-tail enrichment framework for mitigating Gaussian tail independence}

Let $Z\in\mathbb{R}^2$ be the latent Gaussian vector at a given slice, after the AR step, with
\[
E[Z]=0,
\qquad
\operatorname{Cov}(Z)=\Sigma,
\]
where $\Sigma_{11}=\Sigma_{22}=1$ and $\Sigma_{12}=\Sigma_{21}=\rho$.

Let $c\sim\mathcal{N}(0,1)$ be independent of $Z$, and let $\tau\in(0,1)$. Define
\begin{equation}
\alpha=\sqrt{1-\tau^2}
\label{eq:S68}
\end{equation}

Define the direction vector $v\in\{\pm1\}^2$. In the implementation:
\begin{itemize}
    \item if $\rho\geq0$, same-sign corners are used, so $v=[s_x,s_x]^T$;
    \item if $\rho<0$, opposite-sign corners are used, so $v=[s_x,-s_x]^T$,
\end{itemize}
where $s_x=\operatorname{sign}(Z_1)\in\{\pm1\}$. In both cases, each entry satisfies $v_i^2=1$ and $v_1v_2=\operatorname{sign}(\rho)$.

First consider the unconditional case, i.e., the gate is always active. The update is
\begin{equation}
Z'=\alpha Z+\tau c\,v
\label{eq:S69}
\end{equation}

For the mean,
\begin{equation}
\mathbb{E}[Z'] = \alpha \mathbb{E}[Z] + \tau \mathbb{E}[c]\,v = 0
\label{eq:S70}
\end{equation}

For the covariance, using independence between $Z$ and $c$,
\begin{equation}
\operatorname{Cov}(Z')
=
\alpha^2\operatorname{Cov}(Z)
+
\tau^2\operatorname{Cov}(cv)
\label{eq:S71}
\end{equation}

Since $cv$ is a scalar Gaussian multiplied by a fixed vector and $E[c^2]=1$,
\begin{equation}
\operatorname{Cov}(cv)
=
\mathbb{E}\left[(cv)(cv)^T\right]
=
\mathbb{E}[c^2]\,vv^T
=
vv^T
\label{eq:S72}
\end{equation}

Therefore,
\begin{equation}
\operatorname{Cov}(Z')
=
(1-\tau^2)\Sigma+\tau^2 vv^T
\label{eq:73}
\end{equation}

For the diagonal entries,
\begin{equation}
\operatorname{Var}(Z_i')
=
(1-\tau^2)\Sigma_{ii}
+
\tau^2 v_i^2
=
(1-\tau^2)\cdot 1 + \tau^2\cdot 1
=
1
\label{eq:74}
\end{equation}
so the update is exactly variance-preserving for each component. This is the reason for the factor $\sqrt{1-\tau^2}$.

For the off-diagonal term,
\begin{equation}
\operatorname{Cov}(Z_1',Z_2')
=
(1-\tau^2)\rho+\tau^2 v_1v_2
\label{eq:S75}
\end{equation}
and since $v_1v_2=\operatorname{sign}(\rho)$,
\begin{equation}
\rho'
=
(1-\tau^2)\rho+\tau^2\operatorname{sign}(\rho)
\label{eq:S76}
\end{equation}

Thus, the update pushes the correlation toward the intended tail corner direction: toward $+1$ for $\rho>0$ and toward $-1$ for $\rho<0$.

The previous result applies to the unconditional case. In practice, the boosting is applied only in the joint tails, i.e., only when
\begin{equation}
|Z_1|>z_{\mathrm{pivot}},
\qquad
|Z_2|>z_{\mathrm{pivot}},
\qquad
z_{\mathrm{pivot}}=\Phi^{-1}(u_{\mathrm{pivot}})
\label{eq:S77}
\end{equation}

Since the gate is controlled by a high pivot quantile $u_{\mathrm{pivot}}$, it activates only for a small fraction of samples in the joint tails. The coupling is therefore highly localized: it mainly modifies joint extremes while leaving the bulk of the distribution essentially unchanged, especially when $u_{\mathrm{pivot}}$ is high and $\tau$ is moderate.

When the tail condition is satisfied, $Z'$ is nudged toward the empirically observed extreme-corner direction, i.e., same-sign corners for $\rho\geq0$ and opposite-sign corners for $\rho<0$, thereby increasing the frequency of extreme co-occurrences without altering non-tail samples. Because the update is applied conditionally, exact global moment preservation is not guaranteed. However, any deviation in marginal standardization or covariance is typically small due to the rarity of tail events and the bounded injection magnitude.

Finally, the physical marginals remain approximately preserved after mapping through $U=\Phi(Z)$ and the inverse-ECDF transformation, while the frequency of joint extreme events is selectively increased. Even if each $Z_i$ is only approximately standard normal marginally, the tail-gated coupling may introduce small departures from $\mathcal{N}(0,1)$, typically confined to rare joint-tail events. In that case, $U_i=\Phi(Z_i)$ remains bounded in $(0,1)$ but is no longer perfectly uniform, and the inverse-ECDF mapping can slightly perturb the resulting $\theta_i$ marginal, most commonly by allocating slightly more probability to extreme order statistics if the coupling induces heavier tails in $Z$.

In this work, potential tail-induced distortions are controlled by optimizing the pivot quantile $u_{\mathrm{pivot}}$ and coupling strength $\tau$ through an objective loss function that explicitly measures tail agreement between the simulated microstructure and the reference data. This ensures that joint-extreme behavior is improved without degrading the overall statistics.

In summary, the coupling update is exactly variance-preserving when applied unconditionally. When applied under a tail gate, the overall latent distribution becomes a mixture of updated and non-updated samples, so exact global moment preservation is not analytically guaranteed because of tail truncation. However, the gate probability is small for high $u_{\mathrm{pivot}}$, and the injection magnitude is bounded by $\tau$, so any marginal or covariance drift is typically minor. In this work, $u_{\mathrm{pivot}}$ and $\tau$ are selected via Bayesian optimization using a tail-sensitive loss, and the resulting simulated microstructures are verified against the reference data, which implicitly validates that any gated moment deviations remain negligible in the final statistics.

\subsection{Joint-tail enrichment framework for mitigating Gaussian tail independence}

A standard bivariate Gaussian copula exhibits asymptotic tail independence: the conditional probability of one component being extreme given the other is extreme vanishes as the threshold increases.  Real fibre composites, however, display joint-extreme events; both in-plane misalignment angles simultaneously exceeding high thresholds, which a plain Gaussian copula cannot reproduce. The joint-tail enrichment mechanism described below corrects this deficiency at the emission step, without modifying the underlying AR(1) latent chain.

Let $Z_s\in\mathbb{R}^2$ be the clean latent Gaussian vector at a given slice, produced by the AR(1) step. Let $c\sim\mathcal{N}(0,1)$ be independent of $Z_s$, and let $\tau\in(0,1)$ denote the coupling strength.  The coupling scale factor is defined as
\begin{equation}
\alpha=\sqrt{1-\tau^2}.
\label{eq:S75}
\end{equation}
The direction vector $v\in\{\pm1\}^2$ encodes the target corner of the joint tail, determined by the sign of the slice copula correlation: if $\rho\geq0$, same-sign corners are targeted and $v=[s_x,\,s_x]^T$; if $\rho<0$, opposite-sign corners are targeted and $v=[s_x,\,-s_x]^T$, where $s_x=\operatorname{sign}(Z_X)\in\{\pm1\}$.  In both cases $v_i^2=1$ and $v_1v_2=\operatorname{sign}(\rho)$.  The tail-boosted emission vector is then
\begin{equation}
Z_{\mathrm{emit}} = \alpha\,Z_s + \tau\,c\,v.
\label{eq:S79}
\end{equation}

To verify the statistical properties of $Z_{\mathrm{emit}}$, consider first its mean.
Since $\mathbb{E}[Z_s]=\mathbf{0}$ and $\mathbb{E}[c]=0$,
\begin{equation}
\mathbb{E}[Z_{\mathrm{emit}}]
=
\alpha\,\mathbb{E}[Z_s] + \tau\,\mathbb{E}[c]\,v
=
0,
\label{eq:S80}
\end{equation}
so the zero-mean property is exactly preserved.  Turning next to the covariance, and exploiting the independence between $Z$ and $c$,
\begin{equation}
\operatorname{Cov}(Z_{\mathrm{emit}})
=
\alpha^2\,\operatorname{Cov}(Z_s)
+
\tau^2\,\operatorname{Cov}(c\,v).
\label{eq:S81}
\end{equation}
Since $c$ is a scalar Gaussian multiplied by a fixed vector and $\mathbb{E}[c^2]=1$, the second term simplifies as
\begin{equation}
\operatorname{Cov}(c\,v)
=
\mathbb{E}\!\left[(cv)(cv)^T\right]
=
\mathbb{E}[c^2]\,vv^T
=
vv^T,
\label{eq:S82}
\end{equation}
giving
\begin{equation}
\operatorname{Cov}(Z_{\mathrm{emit}})
=
(1-\tau^2)\,\tilde{\Sigma} + \tau^2\,vv^T.
\label{eq:S83}
\end{equation}
Here $\tilde{\Sigma}$ denotes the covariance matrix actually realised by the AR(1) propagation at the current slice: its diagonal entries equal unity exactly (marginal variances are preserved under any convex combination of correlation matrices), while the off-diagonal $\tilde{\rho}=\tilde{\Sigma}_{12}$ is the copula-approximated correlation rather than the ideal target $\rho_s$.

Examining the diagonal entries, note first that $\tilde{\Sigma}_{ii}=1$ exactly: the AR(1) propagation produces $\operatorname{Cov}(Z_s)$ as a convex combination of correlation matrices, and every correlation matrix has unit diagonal, so the diagonal entries are preserved exactly at unity regardless of the transition.  With $v_i^2=1$ by construction,
\begin{equation}
\operatorname{Var}(Z_{\mathrm{emit},i})
=
(1-\tau^2)\,\tilde{\Sigma}_{ii}
+
\tau^2\,v_i^2
=
(1-\tau^2)\cdot 1 + \tau^2\cdot 1
=
1,
\label{eq:S84}
\end{equation}
so each marginal variance is preserved exactly regardless of $\tau$.  This is the design rationale for $\alpha=\sqrt{1-\tau^2}$: it attenuates the clean $Z$ just enough to offset the additional variance injected by the $\tau\,c\,v$ term.  The off-diagonal entry is qualitatively different: $\tilde{\rho}=\tilde{\Sigma}_{12}$ is not an exact quantity but an approximation inherited from the AR(1) propagation (see the preceding subsection), so the emission-layer coupling acts on this already-approximated correlation:
\begin{equation}
\operatorname{Cov}(Z_{\mathrm{emit},1},\,Z_{\mathrm{emit},2})
=
(1-\tau^2)\,\tilde{\rho} + \tau^2\,v_1v_2,
\label{eq:S85}
\end{equation}
and since $v_1v_2=\operatorname{sign}(\tilde{\rho})$, the emission correlation becomes
\begin{equation}
\tilde{\rho}_{\mathrm{emit}}
=
(1-\tau^2)\,\tilde{\rho} + \tau^2\,\operatorname{sign}(\tilde{\rho}).
\label{eq:S86}
\end{equation}
The update pushes $\tilde{\rho}_{\mathrm{emit}}$ toward $+1$ for $\tilde{\rho}>0$ and toward $-1$ for $\tilde{\rho}<0$, selectively increasing the probability of joint-extreme co-occurrences in the corner consistent with the empirical copula sign.

In practice, the enrichment is applied only when both components simultaneously fall in the joint tails,
\begin{equation}
|Z_X|>z_{\mathrm{pivot}}
\quad\text{and}\quad
|Z_Y|>z_{\mathrm{pivot}},
\qquad
z_{\mathrm{pivot}}=\Phi^{-1}(U_{\mathrm{pivot}}),
\label{eq:S87}
\end{equation}
where $U_{\mathrm{pivot}}\in [0.90,1)$ is a high quantile threshold.  Because this gate activates only for a small fraction of draws, the enrichment is highly localised: it modifies joint extremes while leaving the bulk of the distribution essentially unchanged.  When the gate is active, $Z_{\mathrm{emit}}$ is nudged toward the empirically observed extreme-corner direction (same-sign for $\tilde{\rho}\geq0$, opposite-sign for $\tilde{\rho}<0$), increasing the frequency of extreme co-occurrences without altering non-tail samples.  The coupling strength $\tau$ and the pivot quantile $U_{\mathrm{pivot}}$ are both optimised by the Bayesian optimisation procedure using a tail-sensitive composite objective, calibrating joint-extreme behaviour against the reference microstructure.

A fundamental requirement of the implementation is that the tail-boosted vector $Z_{\mathrm{emit}}$ is never fed back as the input to the next AR(1) step.  This design principle is the Clean-latent-state design: where the enhancement is emission-only coupling. As shown in Eq.~\eqref{eq:S76}, the coupling shifts the emission correlation away from $\tilde{\rho}$, the AR(1)-propagated (approximated) latent correlation.  If $Z_{\mathrm{emit}}$ were stored as $Z_{s-1}$ and passed to the next recursion step, the AR(1) propagation would receive a corrupted latent state rather than the theoretically pure $\mathcal{N}(0,\Sigma_{s-1})$ draw required by the recursion.  The covariance invariant $\operatorname{Cov}(Z_s)=\Sigma_s$, derived in Eqs.~(S58)--(S60), would no longer hold, and the corruption would accumulate progressively across all slices of the fibre, degrading the statistical properties of the latent Markov chain.

The clean-Z-state design prevents this by strictly separating latent propagation from emission into two independent operations at each slice:

\begin{enumerate}
    \item {Latent propagation (AR(1) step).}  The clean latent state is advanced
          by one slice:
          \[
          Z_{s-1}^{\,\mathrm{clean}}
          \;\xrightarrow{\;\mathrm{AR(1)}\;}
          Z_s^{\,\mathrm{clean}}
          \;=\;
          \phi_{\mathrm{local}}\,Z_{s-1}^{\,\mathrm{clean}}
          +
          \sqrt{1-\phi_{\mathrm{local}}^2}\,L_s\,\varepsilon_s,
          \qquad \varepsilon_s\sim\mathcal{N}(\mathbf{0},I),
          \]
          giving $Z_s^{\,\mathrm{clean}}\sim\mathcal{N}(\mathbf{0},\tilde{\Sigma}_s)$.  No tail coupling is applied inside this step.

    \item {Emission (tail-enrichment step).}  The coupling transform is applied conditionally at the case of joint-extremes to the clean state to produce the emission vector:
          \[
          Z_{\mathrm{emit}}
          =
          \sqrt{1-\tau^2}\;Z_s^{\,\mathrm{clean}} + \tau\,c\,v,
          \qquad c\sim\mathcal{N}(0,1).
          \]
          $Z_{\mathrm{emit}}$ is used solely to compute the uniform scores $U_i = \Phi(Z_{\mathrm{emit},i})$, which are then mapped to misalignment angles via the empirical inverse-CDF.  It is a transient, slice-local quantity and is discarded immediately after angle emission.

    \item {Chain update.}  The latent chain state committed for the next slice is
          $Z_s^{\,\mathrm{clean}}$, not $Z_{\mathrm{emit}}$:
          \[
          Z_s^{\,\mathrm{chain}} \;\leftarrow\; Z_s^{\,\mathrm{clean}}.
          \]
\end{enumerate}

Under this design, the latent Markov chain $\bigl\{Z_s^{\,\mathrm{clean}}\bigr\}$ evolves according to the theoretically exact AR(1) recursion throughout the full length of each fibre and across the entire microstructure depth, regardless of the tail-enrichment parameters $\tau$ and $U_{\mathrm{pivot}}$.  The tail coupling operates exclusively as an emission-layer transform at each slice, with no feedback path into the latent dynamics.  The AR(1) covariance structure and the joint-tail enrichment mechanism are therefore decoupled by construction: the former guarantees the statistical fidelity of the latent chain, while the latter selectively corrects the joint-extreme behaviour of the emitted angle pairs.

To place these operations in the context of the full generation pipeline, each fibre is synthesised slice by slice as follows.  At every depth slice $s$, the AR(1) step first advances the clean bivariate latent state $(Z_{X,s-1}^{\,\mathrm{clean}},\,Z_{Y,s-1}^{\,\mathrm{clean}})$ to produce $(Z_{X,s}^{\,\mathrm{clean}},\,Z_{Y,s}^{\,\mathrm{clean}})$, preserving the unit-variance Gaussian marginals and the slice-wise copula structure.  The joint tail-enrichment transform is then applied to this clean bivariate pair to yield $Z_{\mathrm{emit}}$, operating on both components simultaneously so that their joint extreme behaviour is steered toward the empirically calibrated corner.  The enriched vector is mapped to the uniform space via $U_i=\Phi(Z_{\mathrm{emit},i})$, and the resulting uniform scores are passed through the per-slice empirical inverse-CDF to recover the physical misalignment angles $(\theta_X,\,\theta_Y)$.  Finally, a local fibre motif may be conditionally overlaid on the physical angle pair, with a pre-defined shape when the motif trigger criterion is met as per to Section 2.3.4 in the main manuscript. Finally, these final misalignment angles are used to map to spatial coordinates.

\newpage

\section{Synthetic microstructure fiber generation: geometry enforcement, neighbor pruning, and the PGS contact solver}

This section explains how the pipeline enforces geometric feasibility, i.e., no overlap, during slice-wise three-dimensional growth. The core difficulty is that candidate fiber centers are proposed statistically, but those proposals can generate local collisions in dense packings. A robust physical synthesis therefore requires: (i) an efficient way to identify which fibers might interact, and (ii) an iterative solver that separates interacting fibers while minimally perturbing the statistically generated configuration.

\subsection{K-candidate selection per fiber (feasibility-aware sampling)}

The statistical generator produces angles that are realistic but may occasionally lead to difficult geometric conflicts in dense regions. To reduce the resampling frequency while maintaining stochasticity, a $K$-candidate mechanism is used: each fiber proposes $K_{\mathrm{select}}$ independent candidates, and the most feasible candidate is chosen.

For each fiber $i$, each candidate draw includes:
\begin{itemize}
    \item an AR(1) latent update, with optional tail coupling,
    \item mapping to uniform variables and slice-wise inverse ECDFs to obtain $\theta_{X,i}$ and $\theta_{Y,i}$,
    \item optional motif blending if motif mode is active,
    \item projection to a provisional center $\mathbf{c}_i^{\mathrm{pre~(s+1)}}$.
\end{itemize}

\subsubsection{Local feasibility scoring (greedy selection)}

Rather than evaluating all pairwise interactions, local feasibility is scored against a neighbor list, i.e., based on Delaunay neighbors from the previous slice. Fibers are processed in a greedy order, typically with larger radii first, since these are the most blocking.

For a candidate of fiber $i$, the gap violations are evaluated only against already fixed neighboring fibers $j$, and a lexicographic score is computed. Specifically, each of the $K$ candidates is projected to its proposed center $\mathbf{c}_i^{\mathrm{cand}}$. It is then checked against every already placed neighbor $j\in N(i)$ that has been fixed earlier in the current greedy pass. For each such neighbor pair, the signed gap is
\begin{equation}
g_{ij}
=
\mathrm{\gamma} ~(r_i+r_j)-d_{ij}
\label{eq:S88}
\end{equation}
where
\[
g_{ij}>0 \quad \text{means overlap (violation)},
\qquad
g_{ij}\leq 0 \quad \text{means separated (acceptable)}.
\]

From all neighbor checks, three scalar statistics are accumulated:
\begin{equation}
N_{\mathrm{viol}}
=
\#\left\{
j : g_{ij}>\varepsilon_{\mathrm{gap}}
\right\}
\label{eq:S89}
\end{equation}
\begin{equation}
\Sigma_{+}
=
\sum_j \max\left(g_{ij}-\varepsilon_{\mathrm{gap}},\,0\right)
\label{eq:S90}
\end{equation}
\begin{equation}
g_{\max}
=
\max g_{ij}
\label{eq:S91}
\end{equation}

The score tuple is then defined as
\begin{equation}
\mathrm{score}
=
\left(
N_{\mathrm{viol}},
\Sigma_{+},
g_{\max}
\right)
\label{eq:S92}
\end{equation}

Lexicographic ordering means that the tuple is compared element by element from left to right. Therefore, the candidate with the lowest score is chosen according to the following priority:
\begin{enumerate}
    \item $N_{\mathrm{viol}}$: the fewest neighbors still violating the gap tolerance,
    \item $\Sigma_{+}$: if there is a tie in the number of violations, the smallest total excess overlap is preferred,
    \item $g_{\max}$: if there is still a tie, the smallest worst-case overlap is selected, so choose $i^* = \arg\min_i \left( \max g_{ij} \right)$
\end{enumerate}

Thus, a candidate with zero violations always takes precedence over one with any violation, regardless of how small that violation may be. Only when two candidates have the same $N_{\mathrm{viol}}$ does the total overlap measure $\Sigma_{+}$ break the tie, and only when these are also equal does $g_{\max}$ determine the final choice.

This scoring strategy is local and greedy: it checks only neighboring fibers $j$ that have already been placed earlier in the greedy ordering during the current round. It is therefore not a global check over all fiber pairs. For this reason, PGS together with full all-pairs verification is still required afterward. The $K$-candidate selection acts only as a bias toward most probable feasible proposals rather than a guarantee of feasibility.

This strategy biases the search toward configurations that are easier to resolve geometrically while preserving randomness, since the candidates originate from independent stochastic draws. Increasing $K_{\mathrm{select}}$ reduces the probability of entering an infeasible or highly colliding state, thereby decreasing the number of rejected rounds and accelerating convergence, without converting the generator into a deterministic optimizer.

\subsection{Efficient spatial interaction graphs for overlap resolution and verification}

A naive approach to overlap resolution and geometric verification would require evaluating all pairwise fiber interactions at each slice, resulting in an $\mathcal{O}(N^2)$ complexity per slice-depth increment, which becomes challenging for fiber counts ranging from several hundred to several thousand. Two distinct spatial data structures are therefore employed to reduce this cost: a Delaunay triangulation, which constructs a sparse interaction graph used exclusively during the iterative PGS solver, and a KDTree, which supports a three-level filtering pipeline used exclusively during the final global verification step. These two mechanisms serve fundamentally different roles and operate at different stages of each round.

\subsubsection{Delaunay interaction graph for the PGS solver}

The PGS solver requires a set of fiber pairs over which the ellipse non-overlap constraints are enforced. Applying constraints to all $\binom{N}{2}$ pairs would be both unnecessary and computationally expensive. Instead, at each geometric fixed-point iteration, a sparse neighbor graph $E_{\mathrm{Del}}$ is constructed from the two-dimensional Delaunay triangulation of the current fiber centers $\{\mathrm{c}_i^{pre}\}_{i=1}^{N}$ at slice $s+1$.

The Delaunay triangulation partitions the point set into non-overlapping triangles such that the circumcircle of each triangle contains no other point in its interior. This maximizes the minimum angle across all triangles and produces the most geometrically regular tessellation possible for a given point set. From the resulting simplices, the undirected edge set is extracted as
\begin{equation}
E_{\mathrm{Del}}
=
\left\{
(i,j)\mid
i \text{ and } j \text{ share an edge in the Delaunay triangulation}
\right\}
\label{eq:S93}
\end{equation}

In practice, each fiber has $\mathcal{O}(1)$ Delaunay neighbors, typically 5 to 7 in two dimensions, so $E_{\mathrm{Del}}=\mathcal{O}(N)$. PGS constraint enforcement is therefore performed exclusively on pairs $(i,j)\in E_{\mathrm{Del}}$, reducing the per-pass cost from $\mathcal{O}(N^2)$ to $\mathcal{O}(N)$.

Critically, the Delaunay graph is rebuilt at every geometric fixed-point iteration, not only once per round. This is necessary because the PGS corrections displace fiber centers, which changes the local spatial arrangement and therefore the set of meaningful geometric contacts. Rebuilding the graph ensures that newly proximate pairs are included in the constraint set and previously separated pairs are no longer processed unnecessarily.

\subsubsection{KDTree-based three-level filtering for global verification}

After the geometrically consistent PGS solver terminates, the proposed fiber configuration at slice $s+1$ must be subjected to an exact global verification step that checks all potentially overlapping pairs, not just those in $E_{\mathrm{Del}}$, before the slice can be accepted. A direct loop over all $\binom{N}{2}$ pairs would again be $\mathcal{O}(N^2)$. Instead, verification uses a KDTree, specifically SciPy's \texttt{cKDTree}, combined with a three-level filter that progressively discards non-interacting pairs at increasing precision.

\paragraph{Tree construction}

A KDTree is a binary space-partitioning tree built on the two-dimensional center coordinates $\{\mathrm{c}_i^{post}\}_{i=1}^{N} \in\mathbb{R}^{N\times 2}$. The tree is constructed by recursively splitting the point set along alternating coordinate axes, first $x$, then $y$, then $x$, and so on, placing the median point at each node. This produces a balanced binary tree in $\mathcal{O}(N\log N)$ time that supports ball-point queries. These queries return all points within a specified Euclidean radius of a query point in $\mathcal{O}(\log N)$ average time per query by pruning entire subtrees whose bounding boxes lie fully outside the query ball. The tree is built once per verification call, after which all $N$ per-fiber queries share it.

\paragraph{Filter level 1: KDTree ball query (conservative enumeration)}

For each fiber $i$, a conservative search radius is computed as
\begin{equation}
r_i^{\mathrm{search}}
=
\gamma\,(a_i+a_{\max})+\varepsilon_{\mathrm{gap}}
\label{eq:S94}
\end{equation}
where $a_i$ is the semi-major axis of ellipse $i$, $a_{\max}=\max_j a_j$ is the global maximum semi-major axis over all fibers, and $\varepsilon_{\mathrm{gap}}$ is the gap tolerance.

The KDTree is then queried for all fiber indices $j$ satisfying $d_{ij}\leq r_i^{\mathrm{search}}$. The use of $a_{\max}$ rather than the specific $a_j$ is deliberate. Since $a_j\leq a_{\max}$ for any $j$, the radius
\[
\gamma\,(a_i+a_{\max})+\varepsilon_{\mathrm{gap}}
\]
is always greater than or equal to
\[
\gamma\,(a_i+a_j)+\varepsilon_{\mathrm{gap}},
\]
guaranteeing that no overlapping pair is ever excluded from the candidate set. Pairs with $j\leq i$ are immediately discarded to avoid double-counting and self-comparisons.

\paragraph{Filter level 2: Bounding-circle cull}

For each candidate pair $(i,j)$ returned by the ball query, a cheaper intermediate test is applied using the actual $a_j$:
\begin{equation}
\text{discard if }
d_{ij}
\geq
\gamma\,(a_i+a_j)+\varepsilon_{\mathrm{gap}}
\label{eq:S95}
\end{equation}

Since $a_i$ is the semi-major axis, i.e., the maximum extent of ellipse $i$ in any direction, a pair whose center-to-center distance exceeds the sum of the bounding-circle radii cannot possibly overlap, regardless of ellipse orientation. This cull eliminates the majority of candidates returned by the ball query at $\mathcal{O}(1)$ cost per pair, before any heavy computation is performed.

\paragraph{Filter level 3: Exact ellipse support-radius check}

For pairs surviving the bounding-circle cull, the exact directional support radii $r_i$ and $r_j$ are evaluated by rotating the separation direction $\mathbf{n}_{ij}$ into each ellipse's local frame and applying the closed-form ellipse boundary formula. The signed gap is then computed as
\begin{equation}
g_{ij}
=
\gamma\,(r_i+r_j)-d_{ij}
\label{eq:S96}
\end{equation}

A positive $g_{ij}$ indicates overlap of magnitude $g_{ij}$, whereas a non-positive $g_{ij}$ indicates separation. The maximum gap across all evaluated pairs is tracked throughout the loop:
\begin{equation}
g_{\max}
=
\max_{i<j} g_{ij}
\label{eq:S97}
\end{equation}

The slice is accepted if and only if $g_{\max}\leq\varepsilon_{\mathrm{gap}}$. If any pair violates this condition, the entire round is rejected without committing any centers, angles, or latent state, and a fresh statistical proposal is drawn for the next round.

This three-level structure, i.e., KDTree ball query, bounding-circle cull, and exact ellipse check, ensures that the computationally expensive ellipse-boundary evaluation is performed only for pairs that genuinely cannot be excluded by cheaper geometric arguments, while maintaining the completeness guarantee that no overlapping pair is ever missed.The 3D growth and overlap solver parameters description is shown in Table~\ref{tab:S1}.

\renewcommand{\arraystretch}{1.2}
\captionsetup[table]{justification=raggedright,singlelinecheck=false}
\begin{longtable}{p{0.16\textwidth} p{0.66\textwidth} p{0.10\textwidth}}
\caption{3D growth and overlap solver parameters, descriptions, and values used in this work.}
\label{tab:S1}\\

\toprule
\textbf{Parameter} & \textbf{Description} & \textbf{This Work} \\
\midrule
\endfirsthead

\toprule
\textbf{Parameter} & \textbf{Description} & \textbf{This Work} \\
\midrule
\endhead

\bottomrule
\endfoot

$K_{\mathrm{select}}$ 
& Number of independent AR(1) statistical proposals generated per fiber per round. The best-scoring one is kept based on the fewest violations and the smallest gaps. Higher values provide a better feasibility bias before PGS, but increase the computational cost per round linearly. 
& 150 \\

$N_{\mathrm{rounds}}$ 
& Maximum number of full round attempts per slice before an error is raised. Failed rounds discard all centers and latent states. 
& 600 \\

$N_{\mathrm{outer}}$ 
& Number of outer fixed-point iterations. Recomputes $\theta^{\mathrm{geom}}$ from the current chord, rebuilds ellipses and Delaunay neighbors, and runs PGS to stabilize the center--geometry dependency. 
& 3 \\

$N_{\mathrm{passes}}$ 
& Number of sequential sweeps over $E_{\mathrm{Del}}$ within one PGS call while the geometry is held fixed. More sweeps provide stronger overlap correction per outer iteration. 
& 12 \\

$\omega$ 
& Relaxation factor for the push-apart correction
$\Delta=\frac{1}{2}\,\omega\,g_{ij}\,\mathbf{n}_{ij}$.
It prevents oscillations in dense packings. Too high may cause oscillations in dense configurations, whereas too low leads to slow convergence. 
& 0.85 \\

$\varepsilon_{\mathrm{pgs}}$ 
& Inner early-stop threshold for PGS sweeps; the solver stops when $g_{\max}<\varepsilon_{\mathrm{pgs}}$. It controls the precision of the inner solve only and does not affect slice acceptance. 
& 0.001 \\

$\varepsilon_{\mathrm{gap}}$ 
& The primary acceptance threshold. A slice is committed only if $g_{\max}\leq\varepsilon_{\mathrm{gap}}$. This sets the maximum permissible penetration between any two ellipses in the final committed configuration. 
& 0.3 \\

$\gamma$ 
& Multiplicative inflation on required separations:
$\gamma(r_i+r_j)-d_{ij}$.
It prevents ellipses from being accepted at exact numerical contact, which is typically ill-conditioned. 
& 1.02 (2\%) \\

$f_{\mathrm{cap}}*$ 
& Caps the ellipse semi-major axis:
$a_i=\min\left(\frac{R_i}{|u_{z,i}|},\,f_{\mathrm{cap}}R_i\right)$.
Without this cap, nearly horizontal fibers, if present, produce unrealistically elongated ellipses that dominate the interaction graph and slow down or destabilize the solver. 
& 12 \\

$\alpha_{\parallel}*$ 
& Caps the fraction of the PGS correction that can act along the fiber's projected tangent direction, i.e., the ellipse major-axis direction in the $XY$ plane. Corrections along this direction are geometrically ineffective, as they slide along the long axis, and can cause lateral drift. The perpendicular component is always applied in full; only the parallel component is clipped. 
& 0.12 (12\%) \\

\end{longtable}
\noindent$^*$\,The PGS overlap-resolution step computes a correction vector $\Delta\mathbf{c}_i$ that pushes fibre $i$ away from its overlapping neighbours.  This vector can be decomposed into a component parallel to the fibre's projected in-plane tangent $\mathbf{t}_i$ and a perpendicular component.  The parallel component is geometrically ineffective at resolving radial overlaps, it only slides the fibre along its own long axis without increasing radial clearance and, if left unconstrained, it accumulates as unrealistic longitudinal drift across slices.  The parameter $\alpha_{\parallel}$ caps the parallel component as a fraction of the total correction magnitude; the full perpendicular component is always applied in its entirety.  The in-plane unit tangent is computed from the local tilt angles as
\begin{equation}
\mathbf{t}_i
=
\frac{
\begin{bmatrix}
\tan\theta_{X,i} ; 
\tan\theta_{Y,i}
\end{bmatrix}
}{
\sqrt{
\tan^2\theta_{X,i}
+
\tan^2\theta_{Y,i}
+\epsilon
}
},
\label{eq:s98}
\end{equation}
where $\epsilon$ is a small constant purely to avoid division by zero. The parallel component of the correction is then clipped so that
\[
\lVert \Delta\mathbf{c}_{\parallel} \rVert
\leq
\alpha_{\parallel}\,
\lVert \Delta\mathbf{c} \rVert .
\]

\medskip

\newpage

\section{Results section}

\subsection{Bayesian optimization optimal hyperparameters}

The Bayesian optimization optimal hyperparameters and weighted loss objective function are reported in Table~\ref{tab:S2}.

\begin{table}[H]
\centering
\caption{Bayesian optimization optimal hyperparameters and weighted loss objective function terms.}
\label{tab:S2}
\renewcommand{\arraystretch}{1.2}
\setlength{\tabcolsep}{12pt}

\begin{tabular}{l c c}
\toprule
\textbf{Item} & \textbf{} & \textbf{Selected Optimal Value} \\
\midrule

Iteration number & $100$ & 36 \\

\midrule
\multicolumn{3}{c}{\textbf{Optimal Hyperparameters}} \\
\midrule

$\phi$ & $[0.90,0.999]$ & 0.922747752993462 \\
$\phi_{\mathrm{Jitter}}$ & $[0,0.01]$ & 0.00035970562260327654 \\
$\tau$ & $[0.100, 0.800]$ & 0.38697692757973057 \\
$U_{\mathrm{pivot}}$ & $[0.900,0.999]$ & 0.9147888250729175 \\

\midrule
\multicolumn{3}{c}{\textbf{Weighted Objective Function Terms}} \\
\midrule

$\mathrm{NRMSE}$ &  & 0.33274 \\
$D_{\mathrm{KS}}$ &  & 0.03657 \\
$\mathrm{TailErr}$ &  & 0.00629 \\
${\left|\Delta\rho\right|}$ &  & 0.0.02575 \\
$\mathrm{JointTail}$ &  & 0.01898 \\
$\mathcal{L}$ &  & 0.12610 \\

\bottomrule
\end{tabular}
\end{table}